\documentclass[fleqn,usenatbib]{mnras}

\usepackage[T1]{fontenc}
\usepackage{ae,aecompl}

\usepackage{graphicx}   
\usepackage{amsmath}    
\usepackage{mathptmx}
\usepackage{txfonts}
\usepackage{xspace}

\newcommand{\units}[1]{~\mathrm{#1}}
\newcommand{\msun}{\mathrm{M}_\odot}
\newcommand{\rvir}{\mathrm{R}_{200}}
\newcommand{\mvir}{\mathrm{M}_{200}}

\newcommand{\auriga}{\textsc{Auriga}\xspace}
\newcommand{\apostle}{\textsc{Eagle}\xspace}
\newcommand{\predestined}{{`predestined'}\xspace}

\renewcommand{\ion}[2]{\hbox{#1\,{\sc #2}}}

\newcommand{\NeVI}{\ion{Ne}{vi}\xspace}
\newcommand{\NeVIII}{\ion{Ne}{viii}\xspace}
\newcommand{\OVI}{\ion{O}{vi}\xspace}
\newcommand{\OVII}{\ion{O}{vii}\xspace}
\newcommand{\OVIII}{\ion{O}{viii}\xspace}
\newcommand{\MgX}{\ion{Mg}{x}\xspace}
\newcommand{\HI}{\ion{H}{i}\xspace}
\newcommand{\HII}{\ion{H}{ii}\xspace}
\newcommand{\HeII}{\ion{He}{ii}\xspace}
\newcommand{\HeIII}{\ion{He}{iii}\xspace}

\usepackage{amsmath} 
\usepackage{gensymb}
\newcommand{\angstrom}{\textup{\AA}}

\title[EAGLE-Auriga: baryon cycle]{EAGLE-Auriga: effects of different subgrid models on the baryon cycle around Milky Way-mass galaxies}
\author[A. J. Kelly et al.]{\parbox[t]{\textwidth}{
Ashley J. Kelly,$^{1}$\thanks{E-mail: a.j.kelly@durham.ac.uk},~
Adrian Jenkins,$^{1}$~
Alis Deason,$^{1, 2}$~
Azadeh Fattahi,$^{1}$~\\
Robert J. J. Grand,$^{3}$~
R{\"u}diger Pakmor,$^{3}$~
Volker Springel,$^{3}$~
Carlos S. Frenk$^{1}$}\vspace{5pt} \\
$^{1}$Institute for Computational Cosmology, Department of Physics, Durham University, Durham DH1 3LE, U.K\\
$^{2}$Centre for Extragalactic Astronomy, Department of Physics, University of Durham, South Road, Durham DH1 3LE, UK\\
$^{3}$Max-Planck-Institut f\"{u}r Astrophysik, Karl-Schwarzschild-Str. 1, 85748 Garching, Germany\\
}
\date{Accepted XXX. Received YYY; in original form ZZZ}

\pubyear{2020}
\begin{document}\label{firstpage}
\pagerange{\pageref{firstpage}--\pageref{lastpage}}
\maketitle

\begin{abstract}
Modern hydrodynamical simulations reproduce many properties of the real universe. These simulations model various physical processes, but many of these are included using `subgrid models' due to resolution limits. Although different subgrid models have been successful in modelling the effects of supernovae (SNe) feedback on galactic properties, it remains unclear if, and by how much, these differing implementations affect observable halo gas properties.
In this work, we use `zoom-in' cosmological initial conditions of two volumes selected to resemble the Local Group (LG) evolved with both the \auriga\ and \apostle\ galaxy formation models.
While the subgrid physics models in both simulations reproduce realistic stellar components of $L^\star$ galaxies, they exhibit different gas properties. Namely, \auriga\ predicts that the Milky Way (MW) is almost baryonically closed, whereas \apostle\ suggests that only half of the expected baryons reside within the halo. Furthermore, \apostle predicts that this baryon deficiency extends to the LG, ($r \leq 1 \units{Mpc}$).
The baryon deficiency in \apostle is likely due to SNe feedback at high redshift, which generates halo-wide outflows, with high covering fractions and radial velocities, which both eject baryons and significantly impede cosmic gas accretion. Conversely, in \auriga, gas accretion is almost unaffected by feedback. 
These differences appear to be the result of the different energy injection methods from SNe to gas. Our results suggest that both quasar absorption lines and fast radio burst dispersion measures could constrain these two regimes with future observations.
\end{abstract}


\section{Introduction}

In the $\Lambda$CDM cosmology, gravitationally bound dark matter structures grow
by a combination of accretion of surrounding matter and mergers with smaller
structures \citep{Frenk:1988}. In this model, galaxies form by the cooling and
condensation of gas at the centres of dark matter haloes
\citep{white:1978,white:1991}. Early tests of these models were carried out
using cosmological simulations including dark matter and baryons
\citep{katz:1991, navarro:1991b, katz:1992, navarro:1993, navarro:1994}. These
simulations were unable to reproduce important properties of real galaxies. In
particular, they produced massive galactic disks that were too compact and
rotated too fast \citep{navarro:1995, navarro:1997}.

These early simulations did not include an efficient injection of energy from
stellar winds and supernova (SN) explosions, a process now commonly referred to
as `feedback'.  Feedback can efficiently suppress star formation (SF) by
ejecting dense, star-forming gas, generating turbulence that disrupts
star-forming regions and driving outflows that eject gas from the interstellar
medium (ISM) in the form of a `hot galactic wind' \citep{matthews:1971,
larson:1974}. Efficient feedback prevents gas from cooling excessively at high
redshift and prematurely turning into stars \citep{white:1978,white:1991,
pearce:1999,sommer-larsen:1999, thacker:2001}.
Efficient feedback, from both SNe and active galactic nuclei (AGN) is now a key
ingredient of modern hydrodynamical simulations. These processes are crucial for
reproducing observed galaxy properties such as the stellar mass function (GSMF),
the mass to size relation and the mass to metallicity relation
\citep[e.g.][]{crain:2009, schaye:2010, lebrun:2014, Vogelsberger:2014a,
schaye:2015, nelson:2019}.

While the inclusion of feedback in simulations is universal, there is no
standard implementation of this process. The complexity of baryon physics,
together with limited resolution, makes it impossible to include feedback
\textit{ab initio} from individual massive stars, SNe or AGN in representative
cosmological simulations.  Instead, simulations rely on `subgrid' prescriptions
of feedback, that is, physically motivated models whose parameters may be
calibrated by reference to observational data. Thus, even though the physical
processes responsible for stellar winds, SNe and AGN feedback are not resolved,
it is hoped that their affects on large scales can be faithfully reproduced.

Fundamentally, the SNe subgrid model describes how SNe energy from a single star
particle, which typically represents a simple stellar population (SSP), is
distributed to neighbouring gas elements.
Energy can be injected into either a single gas resolution element, or into
many, as kinetic \citep{navarro:1993, vecchia:2012} or thermal
\citep{vecchia:2012, schaye:2015} energy, or both
\citep{springel:2003,vogelsberger:2014}. There are subtleties within these
different models such as the amount of energy available per mass of stars
formed, thermal losses, the ratio of thermal to kinetic energy injection, the
decoupling of hydrodynamics to disable cooling and more. Similar considerations
apply to AGN feedback (see \citealt {smith:2018} for an in-depth review).

In modern simulations the free parameters of the SNe and AGN subgrid models are
tuned to reproduce a selection of properties of real galaxies. Gas properties
are rarely included in this calibration and are often taken as model predictions
that can be compared with observational data.
Large-scale gas properties such as cosmic accretion into haloes and onto galaxies
have been studied extensively \citep[e.g.][]{keres:2005, brooks:2009,
oppenheimer:2010, hafen:2019,Hou:2019}. These analyses illustrate how the
injection of gas and metals by feedback complicates the baryon cycle within the
circumgalactic medium (CGM) of galaxies, affecting gas inflow rates onto
galaxies by both reducing the rate of first-time gaseous infall
\citep{voort:2011b, nelson:2015} and by recycling previously ejected winds
\citep{oppenheimer:2010}. However, the sensitivity of these processes to the
details of the subgrid model or the spatial scale at which they are significant
is uncertain \citep{voort:2011a}.


Differences in hydrodynamical solvers introduce further uncertainty in the
cosmological baryon cycle \citep{sijacki:2012, keres:2012, vogelsberger:2012,
torrey:2012, bird:2013, nelson:2013}. In general, it appears that hot gas in
moving-mesh simulations cools more efficiently than in particle-based
simulations; therefore, two simulations with the same subgrid model but
different hydrodynamical solvers can have different gas properties. We do not
investigate the effects of different hydrodynamical solvers in this work,
although we consider the implications in light of our results. As we suspect,
these differences turn out to be secondary to those introduced by the subgrid
models \citep{Hayward:2014,schaller:2015b, hopkins:2018}.

In this paper, we focus on the effects of different implementations of SNe
feedback on the Local Group baryon cycle.  We compare the (untuned) emergent
baryon cycle in the \apostle\ and \auriga\ simulations of two Local Group-like
volumes \citep{sawala:2016, fattahi:2016}. The two simulations use the same
gravity solver and initial conditions but have different subgrid galaxy
formation models, with, in particular, very different approaches to SNe
feedback. The \auriga\ simulations use hydrodynamically decoupled wind particles
that are launched isotropically and, upon recoupling, inject both thermal and
kinetic energy into the surrounding gas. In \apostle, SNe energy is injected as
a `thermal dump' which heats a small number of neighbouring gas elements to
a predefined temperature.

Despite the large differences in the subgrid model, which extends beyond the
implementation of SNe feedback, both of these galaxy formation models produce
galaxies at the present day that match many observed properties. Furthermore,
both models have been demonstrated to be give a good match to properties of the
galaxy population as a whole. The \apostle\ model is the same as that in the
reference \apostle\ simulation \citep{schaye:2015, crain:2015}. The \auriga
model has not been explicitly used in large cosmological simulations; however,
it is based on the model used in the Illustris simulations
\citep{vogelsberger:2013, vogelsberger:2014, torrey:2014}, and is similar to
that in IllustrisTNG \citep{nelson:2018} and Fabel \citep{henden:2018}.

This paper is structured as follows. 
In Section~\ref{sec:the_sample} we introduce our sample of simulated haloes and
describe the stellar properties of the central galaxies, including morphology,
surface density and stellar-mass to halo-mass (SMHM) relation.
In Section~\ref{sec:methods} we detail the SNe subgrid prescriptions and the
tracer particles that facilitate comparisons. We also describe how we calculate
the mock observables for ion column densities and the dispersion measure.
We then present our results, starting with a baryon census around our Local
Group analogues in Section~\ref{sec:baryon_evolution}, and
a particle-by-particle analysis of the `missing baryons' at $z=0$ in
Section~\ref{sec:missing_halo_baryons}.  In Section~\ref{sec:subgrid_diff}, we
attempt to understand how differences in the subgrid models lead to very
different baryon cycles on scales up to $\sim 1 \rm{Mpc}$.
We present predicted observables in Section~\ref{sec:observables}, and discuss
the prospects of constraining the subgrid implementation of SNe feedback from
current and future observational datasets.
Finally, in Section~\ref{sec:conclusions} we discuss the implications of our
results, including several caveats, and summarize our conclusions.

\section{The sample}\label{sec:the_sample}

\begin{table*}
  \caption{Simulation parameters at $z=0$. The columns are: 1) halo
    name in the form, AP-$XX$-$YY$-Ea/Au, where $XX$ and $YY$ identify
    the volume and the halo, respectively; 2) halo virial mass; 3)
    halo radius; 4) baryon fraction within $\rvir$; 
    5) stellar mass within $30\units{kpc}$; 6) the $\kappa_{\rm rot}$
    rotation parameter; 7) radial scale-length; 8) bulge effective
    radius; 9) Sersic index of the bulge.} 
    \label{tab:sim_properties}
    \begin{tabular}{ c | c c | c c c c c c }
        \hline
        & \multicolumn{2}{c}{Dark matter} &\multicolumn{6}{c}{Baryons} \\
        Halo name & $\frac{M_{200}} {[\rm 10^{10} M_{\odot}]}$ & $\frac{R_{200}} {[\rm kpc]}$ & $f_{\rm b}$ & $\frac{M_{*}} {[\rm 10^{10} M_{\odot}]}$ & $\kappa_{\rm rot}$ & $\frac{R_{\rm d}} {[\rm kpc]}$ & $\frac{R_{\rm eff}} {[\rm kpc]}$ & $n$\\
        \hline
        AP-S5-N1-Ea & $0.91$ & $199$ & $0.53$ & $2.03$ & $0.58$ & $3.31$ & $1.00$ & $1.02$ \\
        AP-S5-N1-Au & $0.99$ & $205$ & $0.89$ & $5.89$ & $0.66$ & $1.87$ & $0.55$ & $1.04$ \\
        \hline
        AP-S5-N2-Ea & $0.79$ & $190$ & $0.64$ & $1.60$ & $0.54$ & $3.21$ & $0.91$ & $1.06$ \\
        AP-S5-N2-Au & $0.80$ & $191$ & $0.88$ & $3.95$ & $0.64$ & $2.15$ & $0.49$ & $1.07$ \\
        \hline
        AP-V1-N1-Ea & $1.64$ & $242$ & $0.82$ & $4.49$ & $0.58$ & $3.93$ & $1.05$ & $0.99$ \\
        AP-V1-N1-Au & $1.64$ & $242$ & $0.86$ & $10.51$ & $0.59$ & $7.62$ & $0.68$ & $0.73$ \\
        \hline
        AP-V1-N2-Ea & $1.02$ & $207$ & $0.59$ & $1.62$ & $0.26$ & $3.51$ & $1.21$ & $1.06$ \\
        AP-V1-N2-Au & $1.12$ & $213$ & $0.99$ & $3.77$ & $0.68$ & $3.74$ & $0.89$ & $0.92$ \\
        \hline
    \end{tabular}
\end{table*}

\begin{figure*}
    \centering
    \includegraphics[width=1.00\textwidth]{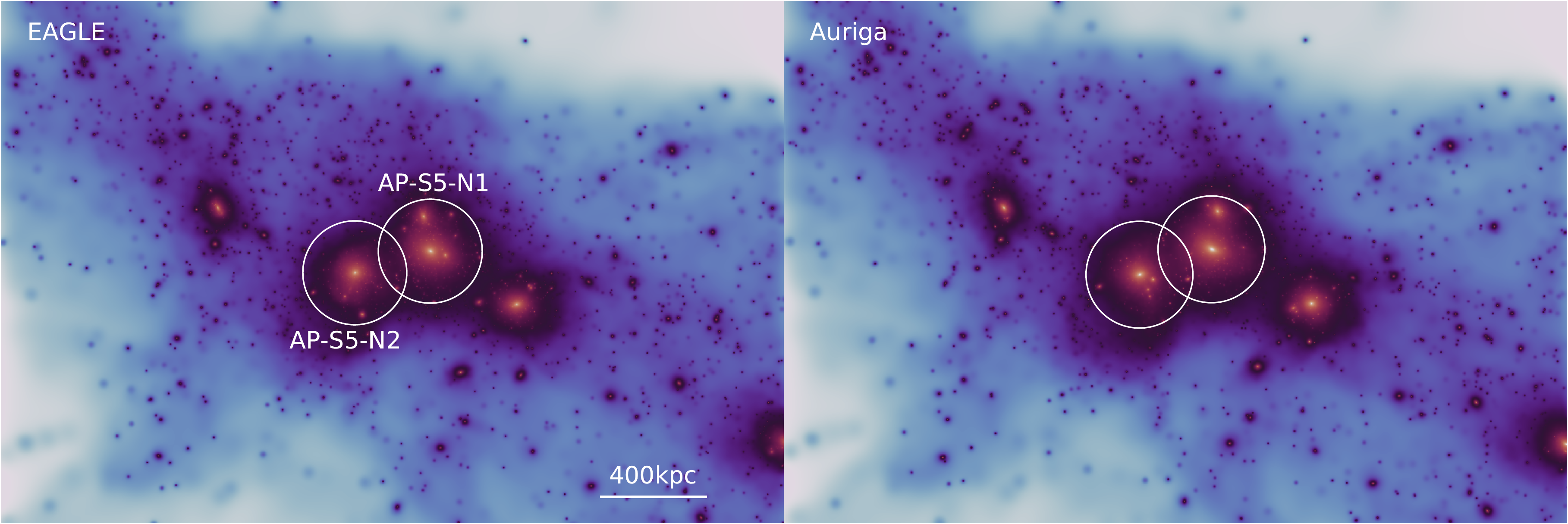}
    \caption{ Present-day mass-weighted density projection of the dark
      matter for the AP-S5 volume in both \apostle\ (left) and
      \auriga\ (right). The projected rectangle has dimensions of
      $(3 \times 2 \times 3) \units{Mpc}$ in the $x$, $y$ and $z$
      directions, respectively, centered on the centre of the mass of
      the two primary haloes. The white circles illustrate $\rvir$ for
      these two haloes. The two main haloes do not overlap in 3D space
      but they do in this particular $x$-$y$ 2D projection.  }
    \label{fig:dm_projection}
\end{figure*}

We focus on $\Lambda$CDM hydrodynamical simulations of two Local Group-like
volumes. Each is a zoom simulation of a region of radius $2-3$~Mpc that contains
a pair of large haloes with virial masses in the range $5 \times 10^{11}
\units{\msun}$ to $2.5 \times 10^{12} \units{\msun}$\footnote{We define the
virial quantities, $\rvir$ and $\mvir$, according to the spherical overdensity
mass \citep{lacey:1994} of each halo centred around the most bound particle
within the halo. $\rvir$ is the radius within which the mean enclosed density,
$\Delta = 200$ times the critical density of the universe.}. We refer to the
four haloes, split across two volumes, as AP-$XX$-N$Y$, where $XX=$S$5$, V$1$
specifies which of the two volumes the halo is in, and $Y=1, 2$ identifies the
two primary haloes. AP-V$1$ and AP-S$5$ correspond to AP-$1$ and AP-$11$ in the
original APOSTLE simulations described in \cite{sawala:2016} and
\cite{fattahi:2016}.
The volumes were selected to match some of the dynamical constraints of the
Local Group. The two primary haloes are required to have present-day physical
separations of $\approx 800 \units{kpc}$ and radial velocities in the range
$(0-250) \units{km~s^{-1}}$. The volumes are also required to have no additional
haloes of mass equal to, or greater than, the least massive of the pair within
a radius $2.5 \units{Mpc}$ of the pair midpoint. More details about the
selection criteria may be found in \cite{fattahi:2016}.

The `zoom-in' initial conditions (ICs) were created using second-order
Lagrangian perturbation theory implemented within \textsc{ic\_gen}
\citep{jenkins:2010}. These ICs have initial gas (dark matter) particle masses
of $1.2 (5.9) \times 10^5 \units{\msun}$, and maximum softening lengths of
$307\units{pc}$. This resolution level corresponds exactly to the L2 resolution in \cite{sawala:2016} and \cite{fattahi:2016}, and is similar to the level 4 resolution in \cite{grand:2017}.

AP-$XX$-$YY$-Ea are the \apostle\ simulations which were run using a highly
modified version of the \textsc{gadget}-3 code \citep{springel:2005}. The fluid
properties are calculated with the particle-based smoothed particle
hydrodynamics (SPH) technique \citep{lucy:1977, gingold:1977}. The \apostle\
simulations adopted a pressure-entropy formulation of SPH \citep{hopkins:2013},
with artificial viscosity and conduction switches \citep{price:2008,
cullen:2010} which, when combined, are known as \textsc{ANARCHY}
\citep{schaye:2015}. 

The \auriga\ simulations, AP-$XX$-$YY$-Au, were performed with the
magnetohydrodynamics code \textsc{arepo} \citep{springel:2010}. The gas is
followed in an unstructured mesh constructed from a Voronoi tessellation of
a set of mesh-generating points which then allow a finite-volume discretisation
of the magneto-hydrodynamic equations. The mesh generating points can move with
the fluid flow. This moving mesh property reduces the flux between cells, thus
reducing the advection errors that afflict fixed mesh codes. For a detailed
description we refer the reader to \cite{springel:2010} and \cite{grand:2017}.

The \auriga\ simulations follow the amplification of cosmic magnetic fields from
a minute primordial seed field. The magnetic fields are dynamically coupled to
the gas through magnetic pressure. \cite{pillepich:2017} demonstrates that the
stellar mass to halo mass (SMHM) relation is sensitive to the inclusion of
magnetic fields, particularly for haloes of $M_{200} \geq 10^{12} \units{\msun}$.
However, this is not important in this work as both galaxy formation models are
calibrated to reproduce realistic $L^\star$ galaxies.

While the general method of calculating the physical fluid properties in the two
simulations is different, there are some similarities. Both numerical schemes
have the property that resolution follows mass, namely, high-density regions are
resolved with more cells or particles. Also, both \apostle\ and \auriga\ have
the same method for calculating gravitational forces: a standard TreePM method
\citep{springel:2005}. This is a hybrid technique that uses a Fast Fourier
Transform method for long-range forces, and a hierarchical octree algorithm for
short-range forces, both with adaptive timestepping.

The initial conditions are chosen to produce present-day Milky Way (MW) and M31
analogues. As both the \auriga\ and \apostle\ simulations share exactly the same
initial conditions, we expect several properties of the simulations to be
similar. Specifically, the dark matter properties should be consistent in both
simulations. Furthermore, as both simulations tune the subgrid models to recover
real galaxy properties, we expect some stellar properties to be similar, but
less so than the dark matter properties.

Dark matter haloes are identified using a Friends-of-Friends (FoF) algorithm
\citep{davis:1985}. The constituent self-bound substructures (subhaloes) within
a FoF group are identified using the \textsc{SUBFIND} algorithm applied to both
dark matter and baryonic particles \citep{springel:2001, dolag:2009}.

Table~\ref{tab:sim_properties} lists properties of the two primary haloes in both
volumes. We see that the baryonic properties of the four haloes in the two
simulations differ somewhat, with up to a factor of two difference in the
stellar mass. We also tabulate the baryon fraction, $f_{\mathrm{b}}$, in each
halo, which we define as the ratio of baryonic to total mass normalised by the
mean cosmic baryon fraction, $\Omega_{\rm b}/\Omega_{\rm m} \sim 0.167$, within
$\rvir$. We find that the baryon fraction in the \apostle\ simulations is
systematically lower than in their \auriga\ counterparts. The virial properties
of the haloes are all consistent, however, with small differences in virial mass
and radius which are due to the different halo baryon fractions.

Fig.~\ref{fig:dm_projection} shows a dark matter projection of the AP-S$5$
volume in both \apostle\ and \auriga. A visual inspection shows that the dark
matter distribution of the two haloes, and their local environment, are almost
identical in the two simulations. There is some variation in the location of
satellite galaxies and nearby dwarf galaxies; this is likely due to the
different baryonic properties and the stochastic nature of N-body simulations.

\begin{figure}
    \centering
    \includegraphics[width=\linewidth]{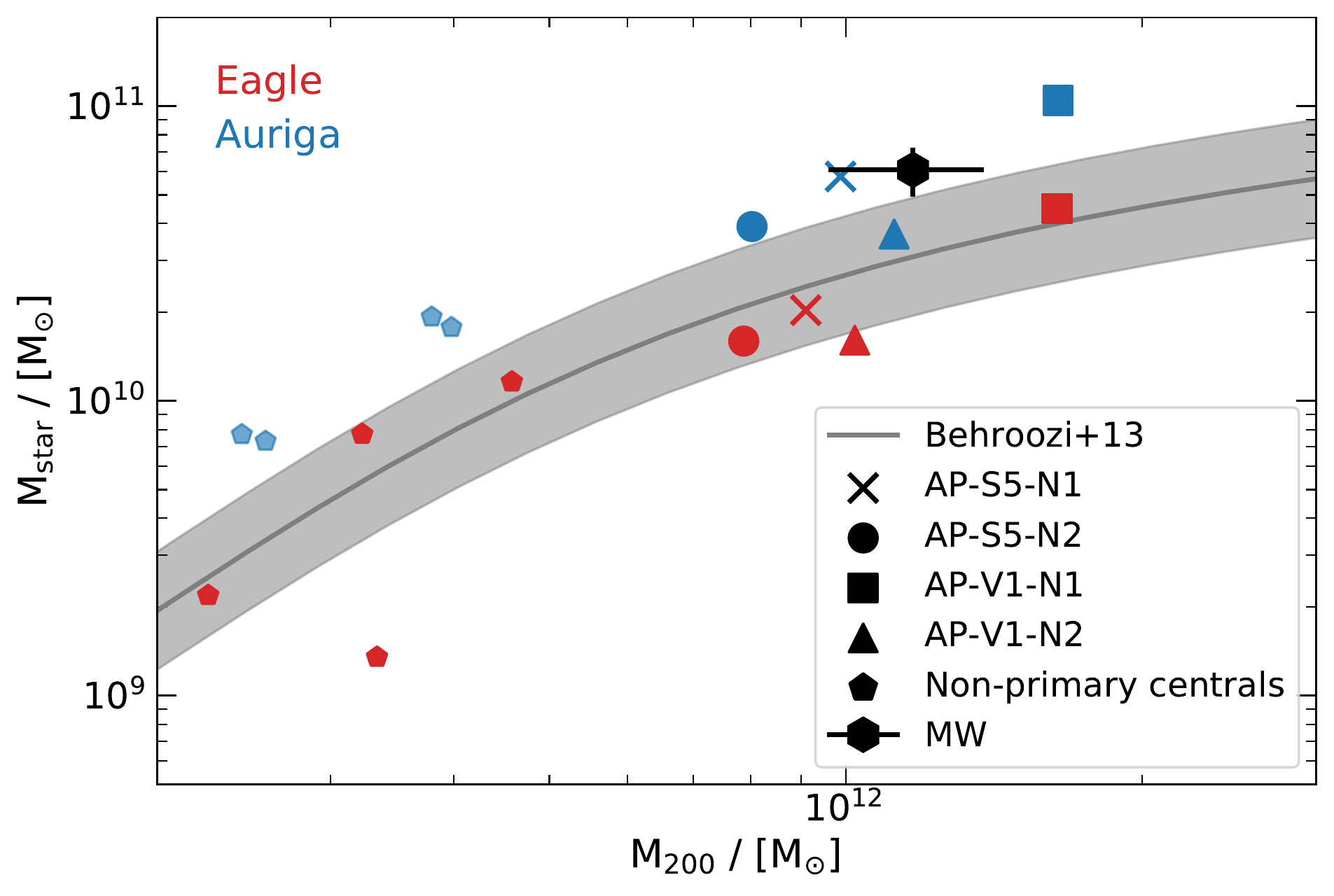}
    \caption{The stellar-mass-to-halo-mass relation in the \apostle\
      (red) and \auriga\ (blue) simulations. The four different
      symbols (cross, circle, square and triangle) differentiate the
      four different primary haloes. The smaller blue and red pentagons
      show the SMHM relation for `central' field galaxies in the
      \apostle\ and \auriga\ simulations respectively. The black solid
      line and grey shaded region are the results of 
      \protect\cite{behroozi:2013} with an estimated scatter of
      $\pm 0.2 \units{dex}$.  We also show as a black hexagon with
      errorbars the estimate for the Milky Way halo mass by
      \protect\cite{callingham:2019} and of the stellar mass by
      \protect\cite{licquia:2015} }
    \label{fig:stellar_halo}
\end{figure}

In Fig.~\ref{fig:stellar_halo} we plot the SMHM relation for the four primary
haloes in the two simulation volumes, \apostle\ and \auriga, alongside the
inferred relation from abundance matching \citep{behroozi:2013}. We also include
several resolved lower mass field `central'\footnote{`Central' refers to the
most massive subhalo with a FoF \citep{davis:1985} group.} haloes from both
\apostle\ and \auriga\ to give an indication of the SMHM over a broader mass
range. We define the stellar masses as the total stellar mass within an aperture
of $30 \units{kpc}$.
The stellar masses of the \auriga\ galaxies are consistently larger than those
of the \apostle\ galaxies, at a given halo mass, over the entire range of halo
masses. As both \apostle\ and \auriga\ model the same volume, the differences
are not due to sample variance.

We also briefly analyse the stellar surface density profiles of the four primary
haloes at the present day. We fit the surface density profiles with a combination
of an exponential profile of scale radius, $R_{\rm{D}}$, and a Sersic profile of
the form $\exp (R/R_{\rm{eff}})^n$ \citep{sersic:1963}. The values of the
best-fit parameters, $R_{\rm{D}}$, $R_{\rm{eff}}$ and $n$, are given in
Table~\ref{tab:sim_properties}. The fit parameters of the models are
consistent with the isolated MW-mass galaxies, the original \auriga\ haloes,
presented in \cite{grand:2017}. Furthermore, the stellar surface density at the solar
radius is a few times $\sim 10 \units{\msun~pc^{-2}}$ in all of the haloes which is consistent
with estimates for the Milky Way \citep{flynn:2006}. The surface density
profiles, and best fit models, can be seen in Fig. \ref{fig:surface_den}. The
galaxy stellar surface density profiles are similar in most cases in the two
simulations, albeit with a systematically higher surface density in the case of
\auriga. 

\begin{figure*}
    \centering
    \includegraphics[width=\linewidth]{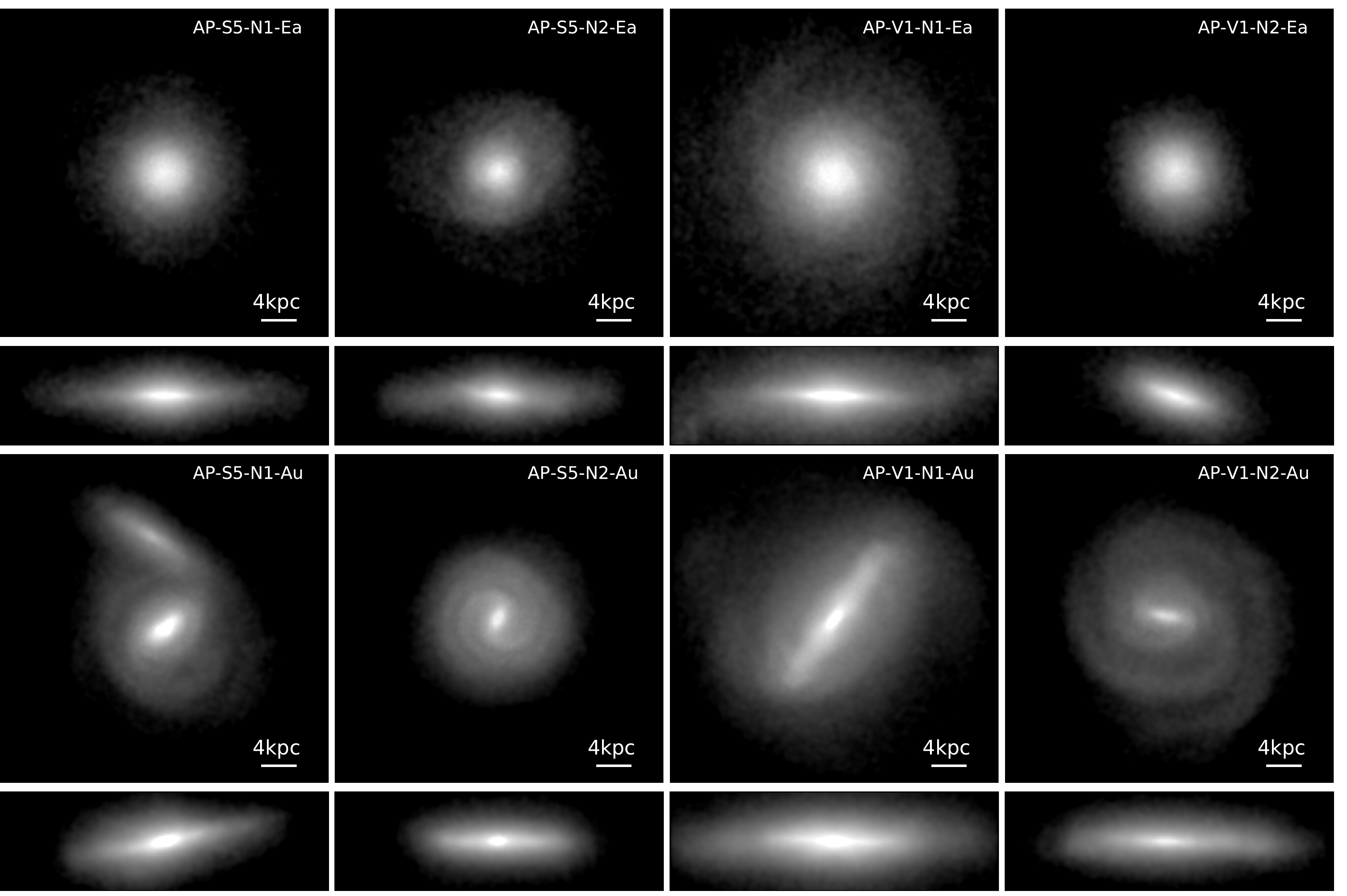}
    \caption{Mass-weighted face-on (top) and edge-on (bottom)
      projections of the stars in the four haloes in the \apostle\ and
      \auriga\ simulations at $z=0$}

    \label{fig:stellar_proj}
\end{figure*}

We also calculate the $\kappa_{\mathrm{rot}}$ rotation parameter for each
galaxy, a measure of the fraction of kinetic energy in organized rotation, which
correlates with morphology \citep{sales:2012}. The quantity
$\kappa_{\mathrm{rot}}$ is defined such that $\kappa_{\mathrm{rot}} \approx 1$
for discs with perfect circular motions and $\kappa_{\mathrm{rot}} \approx 1/3$
for systems with an isotropic velocity dispersion. Thus, a large
$\kappa_{\mathrm{rot}}$ indicates a disc galaxy, whereas a lower value indicates
an elliptical galaxy.
$\kappa_{\mathrm{rot}}$ requires a definition of the z-axis which we take to be
the direction of the total angular momentum of all stars within $30\units{kpc}$
of the centre of the galaxy.
The large values of $\kappa_{\mathrm{rot}}$ in Table~\ref{tab:sim_properties}
are consistent with a visual inspection of Fig.~\ref{fig:stellar_proj}, which
shows face-on and edge-on stellar projections of the four primary haloes. In
Fig.~\ref{fig:stellar_proj} we see that most of the galaxies appear to be
`disky' in projection, with the exception of AP-V$1$-N$2$-Ea, which has a low
$\kappa_{\mathrm{rot}}$. 

In general, the \auriga\ simulations exhibit more recognisable morphological
features, spiral arms and distinct bar components (AP-S$5$-N$2$, AP-V$1$-N$1$
and AP-V$1$-N$2$). The \apostle\ simulations produce smoother looking discs,
with little morphological evidence of either a bar or spiral arms. It is unclear
what causes these morphological differences, but they could be due to different
effective spatial resolutions in the two simulations. In SPH simulations, the
smoothing length determines the spatial resolution, and the ratio of the
smoothing length to the mean inter-particle separation is usually a free
parameter taken to be $\sim 1.3$ \citep{price:2012}. Whereas in moving mesh
simulations, the cells have a spatial radius of approximately $\sim 0.5$ the mean
cell separation. Thus moving mesh simulations typically have a $\sim1.3/0.5\sim2.4$
times better spatial resolution at the same mass resolution\footnote{By the same
mass resolution, we mean that the cell mass in a moving mesh simulation is equal
to the particle mass in an SPH simulation, as in this work.}.

\section{Methods and observables}\label{sec:methods}

\subsection{Subgrid physics models}\label{sec:subgrid}

\apostle\ and \auriga\ include prescriptions for subresolution baryonic
processes such as star formation \citep{schaye:2008}, metal enrichment
\citep{wiersma:2009b}, black hole seeding and growth, active galactic nuclei
(AGN) feedback \citep{springel:2005, booth:2009, guevara:2015}, radiative cooling
\citep{wiersma:2009} and feedback from stellar evolution \citep{vecchia:2012}.
However, as previously noted, the implementations are rather different. In this
section, we describe the qualitative differences in the SNe feedback
prescriptions in the two models.

Traditionally, the energy from SN events occurring within the single stellar
population (SSP) represented by a star particle is injected into a large mass of
local gas \citep{vecchia:2008}. If SNe energy from an
SSP is injected over a mass of gas comparable to the initial stellar mass
formed, the gas is heated to high temperatures, $T \geq 10^{7} \units{K}$.
However, when the same amount of energy is distributed over a much larger mass
of gas, the temperature increase experienced by the gas is much lower. This
lower post-SNe gas temperature results in a shorter cooling time. When the
cooling time is significantly shorter than the sound-crossing time of the gas,
energy injection from SNe is unable to drive a galactic wind efficiently
\citep[e.g.][]{vecchia:2012}.

In the \apostle\ simulations all the SNe energy from an SSP is injected in the
form of thermal energy \citep{schaye:2015}. Rather than distributing the energy
evenly over all of the neighbouring gas particles, the energy is injected into
a small number of neighbours \textit{stochastically} \citep{vecchia:2012}. This
method allows the energy per unit mass, which corresponds to the temperature
change of a gas particle, to be defined. In these simulations each gas particle
heated by SN feedback is always subject to the same temperature increase,
$\Delta T_{\rm{SN}} = 10^{7.5} \units{K}$.

The SNe feedback scheme in the \auriga\ simulations consists of an initially
decoupled wind whose main free parameters are the energy available per unit mass
of SNII and the wind velocity. The wind velocity scales with the 1-dimensional
velocity dispersion of local dark matter particles. Qualitatively, SNe winds are
modelled by `wind particles' which are launched in an isotropic random direction
carrying mass, energy, momentum and metals. Upon launch, the wind particles are
decoupled from hydrodynamic forces and experience only gravity. The wind
particles can recouple either when they reach a region of low-density gas (5\%
of the star-formation density threshold) or when they exceed a maximum travel
time ($0.025$ of the Hubble time at launch). When the wind recouples, it
deposits energy, momentum and metals into the gas cells it intersects.

We do not describe the AGN feedback models but detailed explanations of them can
be found in \cite{schaye:2015} and \cite{grand:2017} for \apostle\ and \auriga,
respectively. The AGN model used in the \apostle\ simulations in this work
differs slightly from the reference model in \cite{schaye:2015}. Namely, in this
work the AGN model uses a viscosity which is hundred times lower than the
reference model this reduces the accretion rate and growth of the black holes.
This model is referred to as `ViscLo` in \cite{crain:2015}. In Section
\ref{sec:conclusions} we discuss the possible effects of the different AGN
models on our results.


\subsection{Tracer particles and particle matching}
\label{sec:tracers_and_pmatching}

The quasi-Lagrangian technique of \textsc{arepo} allows mass to advect between
gas cells so each cell may not represent the same material over the course of
the simulation. The \auriga\ simulations, however, include Lagrangian Monte
Carlo tracer particles \citep{genel:2013, defelippis:2017} which enable us to
track the evolutionary history of individual gas mass elements in a way that
allows direct comparison to SPH gas particles in \apostle. The Monte Carlo
tracer particles have been shown to reproduce the density field in various
tests, including cosmological simulations, accurately \citep{genel:2013}.

In \auriga, a single tracer particle is attached to each gas cell at the
beginning of the simulation. As the simulation proceeds, tracer particles can
transfer across cell boundaries with a probability given by the ratio of the
outward-moving mass flux across a face and the mass of the cell. This allows the
tracer particles to emulate the evolution of a Lagrangian gas element.  The
tracer particles do not carry any physical properties. Instead, they inherit the
properties of the baryonic element to which they are attached at any given time.
The tracer particles introduce a Poisson noise due to their probabilistic
evolution. However, as we use several million tracer particles, this noise is
insignificant.


A combination of identical initial conditions and the tracer particles in
\auriga\ allows us to perform a detailed comparison between the two simulations
on the scales of individual baryonic mass-elements. In the \apostle\
simulations, each dark matter particle in the initial conditions is assigned
a gas particle at the start of the simulation. This represents the baryonic mass
from the same Lagrangian region as the associated dark matter particle.
Likewise, the dark matter particles within \auriga\ are assigned tracer
particles. Both the tracer particles in \auriga, and the gas particles in
\apostle, are assigned permanent, unique ID's dependent on their parent dark
matter particle. The unique ID assigned to each particle facilitates direct
comparison of the same baryonic mass elements between the two simulations.

\subsection{Ion number densities}\label{sec:calc_ion_density}

We calculate column densities of several ionised species following
\cite{wijers:2019}. The total number of ions, $N_{X_{i}}$, of each species in
a given mass of gas is given by, 
\begin{equation}
    N_{X_{i}} = \frac{m_{X} f_{X_{i}}}{m_{Z}},
    \label{eq:ion_number}
\end{equation}
where $m_{X}$ is the total mass of element $X$, $f_{X_{i}}$ is the ionization
fraction of the $i^{th}$ ion, $m_{Z}$ is the mass of an atom of element $X$ and
$Z$ is the atomic number of the ionized element.

We calculate the ionisation fraction of each species using the lookup tables of
\cite{hummels:2017}. These are computed under the assumption of collisional
ionisation equilibrium (CIE). They only consider radiation from the metagalactic
UV background according to the model of \cite{haardt:2012} in which the
radiation field is only a function of redshift. The lookup tables are generated
from a series of single-zone simulations with the photoionisation code,
\textsc{cloudy} \citep{ferland:2013} and the method used for the ``grackle''
chemistry and cooling library \citep{smith:2016}. The ion fractions are
tabulated as a function of log temperature, log atomic hydrogen number density
and redshift.

In both \apostle\ and \auriga\ the masses of some elements are tracked within
the code; these are hydrogen, helium, carbon, nitrogen, oxygen, neon, magnesium,
silicon and iron. We calculate the number of species of each element from
Eq.~\ref{eq:ion_number} with the total mass of each element as tracked by the
code, and the calculated mass fraction of each species state. We can then make
2-dimensional column density maps of each species by smoothing the contribution
from each gas particle onto a 2D grid with a two-dimensional SPH smoothing
kernel. We make these 2D maps by projecting through each of the $x$, $y$ and $z$
axis. We then separate the 2D maps, which cover a region of $(8 \times
8)~\rvir$, into $100$ annuli with linearly spaced radii. In each annulus, we
have many lines-of-sight through the halo. We then take the column density at
radius, $r$, to be the median column density calculated within the annulus
that encloses the radius, $r$, for all three projected maps. We choose to use
the median as it is more representative of a single random line-of-sight
through a halo in the real universe. However, the mean value, which is often
significantly higher, is also of interest as observed values can be biased by
the detection thresholds of instruments.

\apostle\ and \auriga\ use different yield tables when calculating the fraction
of different elements returned from SNIa to the ISM. These yield tables
typically differ by $<20\%$ for the species considered in this work. However,
these difference are negligible in the results presented in Section
\ref{sec:ions_obs}, which span many orders of magnitude. It is possible to
normalise these yield fractions in post-processing, but we choose to use the
simulation tracked quantities are they are self-consistent with the gas cooling.

\begin{figure*}
    \centering
    \includegraphics[width=\linewidth]{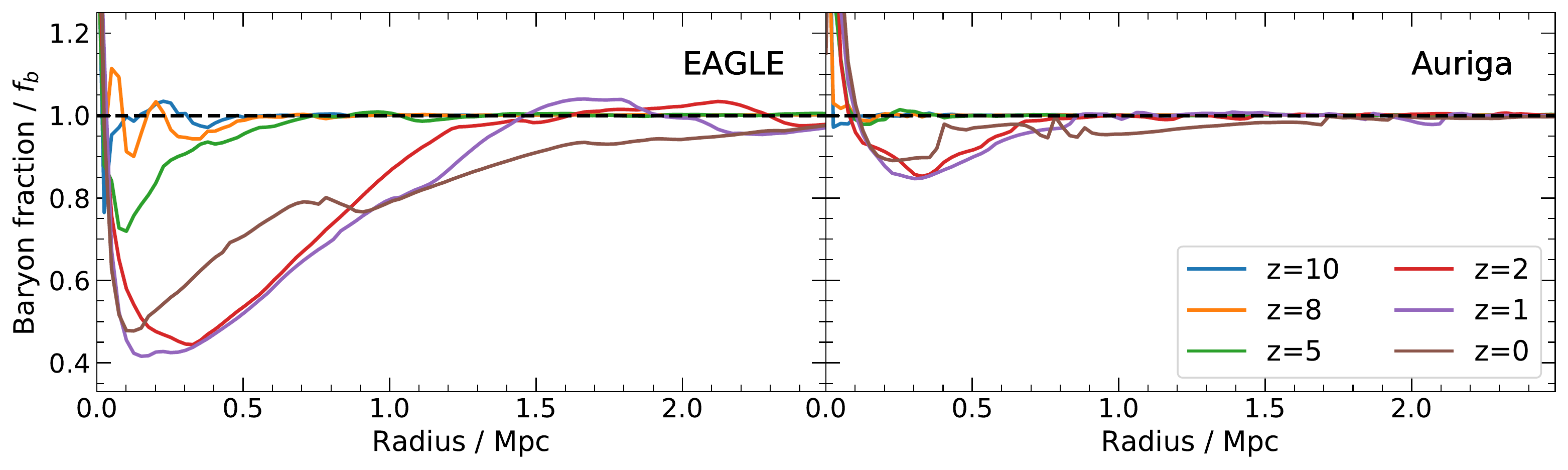}
    \caption{The baryon fraction within a sphere centred around the
      primary halo, AP-S$5$-N$1$, as a function of the sphere
      radius. The baryon fraction is calculated within $100$ spheres
      with a linearly increasing radius in the range
      $0-2.5 \units{Mpc}$ (comoving). We repeat the process at six
      redshifts, $z=0,~1,~2,~5,~8$ and $10$, which are shown by the
      brown, purple, red, green, orange and blue solid lines,
      respectively. The left panel shows the results in \apostle\ and
      the right panel in \auriga. The baryon fraction is normalised
      by the mean baryonic-to-dark matter ratio, $f_b$, in the
      universe. The dashed black line indicates a baryonically closed
      system.}
    \label{fig:baryon_radius}
\end{figure*}

\begin{figure}
    \centering
    \includegraphics[width=1.0\linewidth]{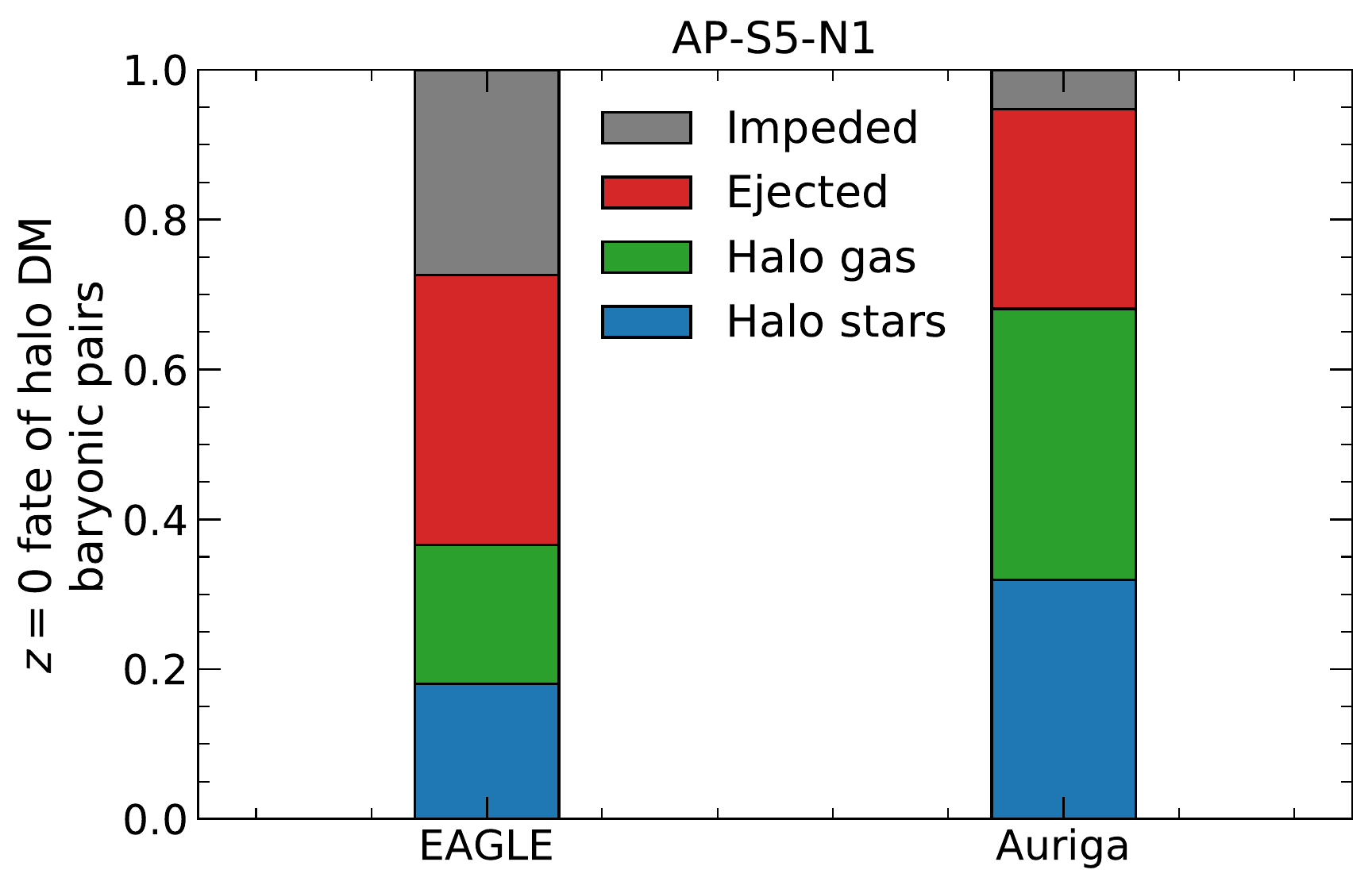}
    \caption{The present-day fate of the baryonic counterparts of the
      dark matter particles within $\rvir$ of the
      primary halo AP-S$5$-N$1$ in \apostle\ and \auriga. The halo
      stars, halo gas, ejected gas and impeded gas are shown by blue,
      green, red and grey bars, respectively. 
    See the text for a detailed description of these categories.}
    \label{fig:missing_halo_baryons_categorised}
\end{figure}

\subsection{The dispersion measure}\label{sec:calc_dm}

Fast radio bursts (FRBs) are bright pulses of radio emission with periods of
order milliseconds, typically originating from extragalactic sources \citep[see
review by][]{cordes:2019}. The first FRB, which was reported by
\cite{lorimer:2007}, was found in archival data from the $64 \units{m}$ Parkes radio
telescope. By 2019, there were over 80 distinct FRB sources reported in the
literature \citep[see review by][]{petroff:2017}. In the next few years, these
FRB catalogues will grow by orders of magnitude with current and future surveys
detecting thousands of events per year \citep{connor:2016}. 

As radiation from an FRB propagates through the intervening gas, the free
electrons in the gas retard the radiation. As the retardation of the radiation
is frequency dependent, this process disperses the FRB pulse, thus producing
a measurable time delay between the highest and lowest radio frequencies of the
pulse. The dispersion measure quantifies this time delay. The dispersion
measure, from observations of the photon arrival time as a function of
frequency, is given by, 
\begin{equation}
    \int (1+z)~n_{e}~dr~,
    \label{eq:dm}
\end{equation}
which provides an integral constraint on the free electron density, $n_{e}$,
along the line-of-sight from the observer to the source, where $dr$ is the path
length in comoving coordinates. The free electron density and the radiation path
in Eq.~\ref{eq:dm} are also in comoving units. The dispersion measure will
include contributions from free electrons in the IGM \citep{zheng:2014,
shull:2018}, our galaxy, the Local Group, the galaxy hosting the FRB and baryons
residing in other galactic haloes which intersect the sightline
\citep{mcquinn:2014}.  As such, FRBs provide a possible way to investigate the
presence of baryons that are difficult to observe with other methods. 

In hydrodynamical simulations, we can calculate contributions to the dispersion
measure from both the ISM and the hot halo of MW-like galaxies, and investigate
the model dependence. The electron column density can be calculated for
sightlines in a similar way to that described in Section~\ref{sec:ions_obs}
below. We calculate the number of free electrons for each gaseous particle or
cell in the simulations by computing the number density of \HII, \HeII\ and
\HeIII, which dominate the total electron density; these calculations again
utilise the ion fraction lookup tables of \cite{hummels:2017}. 

\section{Baryon evolution}\label{sec:baryon_evolution}

As we have shown in Section~\ref{sec:the_sample}, \apostle\ and \auriga\ produce
galaxies in MW-mass haloes that have roughly similar morphologies, stellar masses
and stellar mass distributions. However, as the simulations are calibrated to
reproduce a number of stellar properties of the observed galaxy population,
these similarities are not too surprising. In this section we explore the
effects of the two different hydrodynamical schemes and feedback implementations
on the \textit{untuned} baryon properties. In particular, we investigate the
baryon fraction around the two pairs of MW and M31 analogues and how this
evolves with both radius and time.

In Fig.~\ref{fig:baryon_radius} we investigate the baryon fraction within
a sphere centred on the main progenitors of the AP-S$5$-N$1$ simulations, as
a function of radius at six different redshifts. Even though both simulations
follow the assembly of the same dark matter halo, the baryon fraction within
a sphere of radius $\approx 500 \units{kpc}$ (comoving) begins to differ
significantly between $z=8$ and $z=5$. By $z=2$ the \apostle\ model has
developed a baryon deficiency of $\geq 10\%$ within a radius of $\geq
1 \units{Mpc}$ (comoving), extending to $\approx 2 \units{Mpc}$ at the present
day. We refer to this is the Local Group baryon deficiency.
By contrast, in the \auriga\ simulations the baryon fraction is within $10\%$ of
unity for radii $\geq 0.5 \units{Mpc}$ (comoving) at all redshifts. Furthermore,
the minimum baryon fraction within a sphere around the  primary \auriga\ galaxy
is $\approx 80\%$, approximately twice that of its \apostle\ counterpart. 

In Table~\ref{tab:sim_properties} the baryon fraction of AP-V$1$-N$1$-Ea is
similar ($\sim 0.85\Omega_{\rm b}/\Omega_{\rm m}$) to that of the \auriga\
counterpart. However, the baryon fraction of this galaxy at $z=1$ was $\sim 0.5
\Omega_{\rm b}/\Omega_{\rm m}$, almost a factor of two lower than in \auriga.
Furthermore, the baryon fraction within a radius of $\sim 1 \units{Mpc}$ is
$\sim 30\%$ lower than in \auriga. Thus, even the most baryon rich halo in
\apostle\ is still baryon poor compared to the same halo in \auriga. The
differences in the baryon fraction of the haloes in the local region, out to
$\sim 1~ {\rm Mpc}$,  around AP-S$5$-N$1$ are representative of the sample.
Thus, while we focus on the individual halo AP-S$5$-N$1$ for illustration, the
general results are valid for all haloes in our sample.

In both \apostle\ and \auriga\ the halo baryon deficiency peaks at around $z=1
- 2$, which is consistent with the observed peak in the star formation rate in
the real universe \citep{madau:2014}. However, the amplitude, spatial extent and
scale of the baryon deficiency in \apostle, compared to \auriga, is striking.
This difference is particularly remarkable given that the primary galaxies are
relatively similar.

AGN feedback is often thought to be the cause of low baryon fractions in
MW-mass, and more massive haloes in cosmological hydrodynamical simulations
\citep{bower:2017, nelson:2018, davies:2019}. Nevertheless, we see from
Fig.~\ref{fig:baryon_radius} that the decrease in baryon fraction in \apostle\
sets in at high redshift when the halo masses, and thus the black hole masses,
are much lower. These results thus imply a different driver for the low baryon
fractions at the present day.

Although the two simulations predict very different local baryon fractions, both
within the halo and beyond, it is difficult to ascertain which of these models
is the more \textit{realistic}. In practice, it is very difficult to infer the
baryon fractions of real galaxies. For external galaxies this are typically
derived from X-ray emission using surface brightness maps to infer a gas density
profile. These inferences also require information about the temperature and
metallicity profiles, which are difficult to measure accurately with current
instruments \citep{bregman:2018}.

In the Milky Way, the baryonic mass in stars, cold and mildly photoionised gas
and dust is estimated to be $0.65 \times 10^{11} \units{\msun}$
\citep{McMillan:2011}. The total mass of the halo, which is dominated by dark
matter, is estimated to be $(1.17 \pm 0.15) \times 10^{12} \units{\msun}$
\citep{callingham:2019}. Assuming a universal baryon fraction of $\Omega_{\rm
b}/\Omega_{\rm m} \approx 0.157$ \citep{planck:2013} implies that $0.8 \times
10^{11} \units{\msun}$ of the MW's baryons are unaccounted for. There is strong
evidence that some fraction of these unaccounted baryons resides in a hot
gaseous corona surrounding massive haloes; however, it total mass remains
uncertain.  An accurate estimate of the mass of the MW hot halo requires
knowledge of the gas density profile as a function of radius. While the column
density distribution of electrons is constrained in the direction towards the
LMC \citep{anderson:2010}, there is a degeneracy between the normalisation and
the slope of the electron density profile \citep{bregman:2018}. Assuming a flat
density profile \citep[e.g.][]{gupta:2012, gupta:2014, faerman:2017}, the MW
appears to be `baryonically closed'. However, for faster declining density
profiles \citep[see][]{miller:2015, li:2017b}, the gaseous halo out to $\rvir$
contains only half of the unaccounted baryons. Therefore, the predictions of
Fig.~\ref{fig:baryon_radius} for both \apostle\ and \auriga\ are consistent with
observations given the uncertainties.

\section{The missing halo baryons}\label{sec:missing_halo_baryons}

\begin{figure*}
    \centering
    \includegraphics[width=0.85\textwidth]{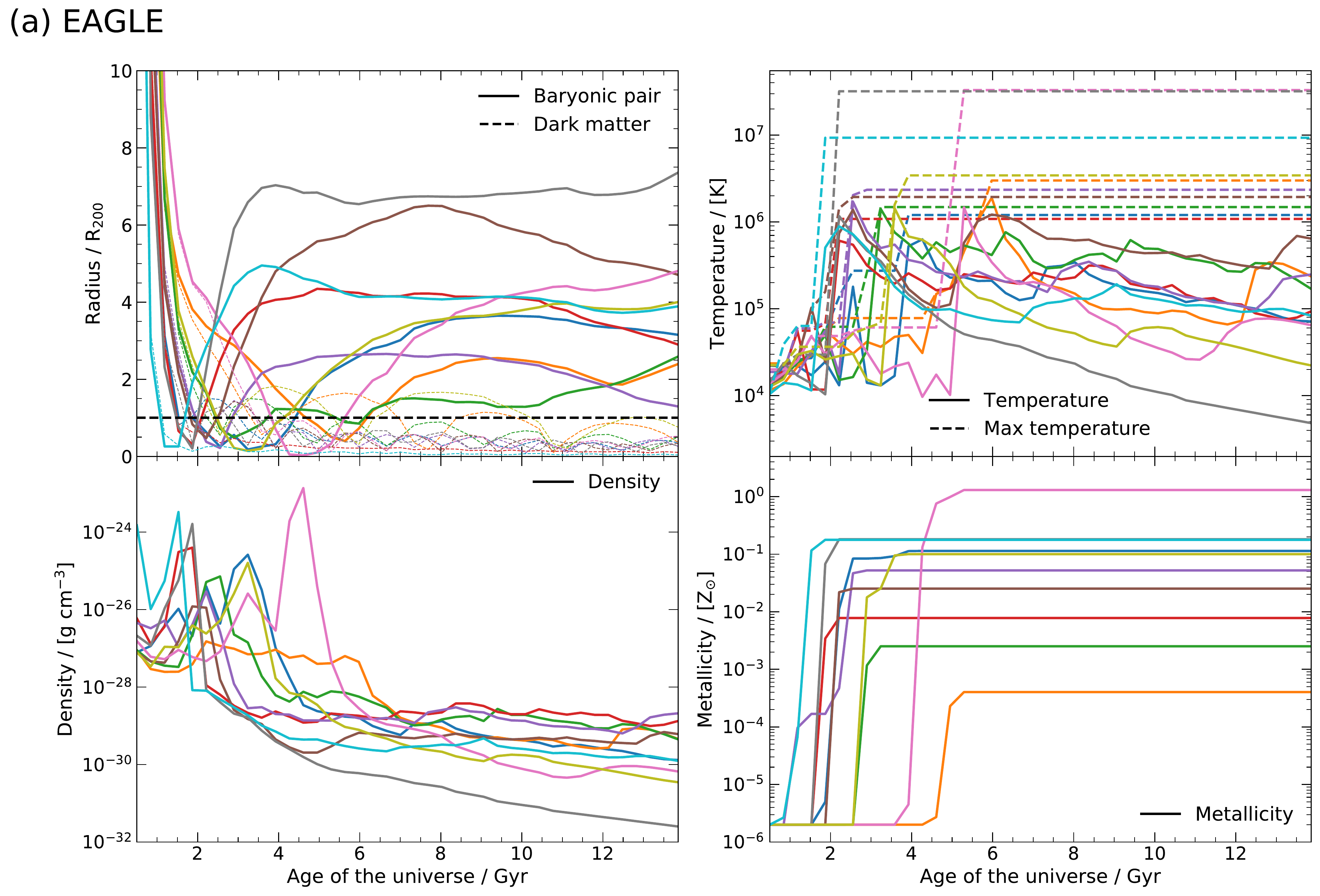}
    \includegraphics[width=0.85\textwidth]{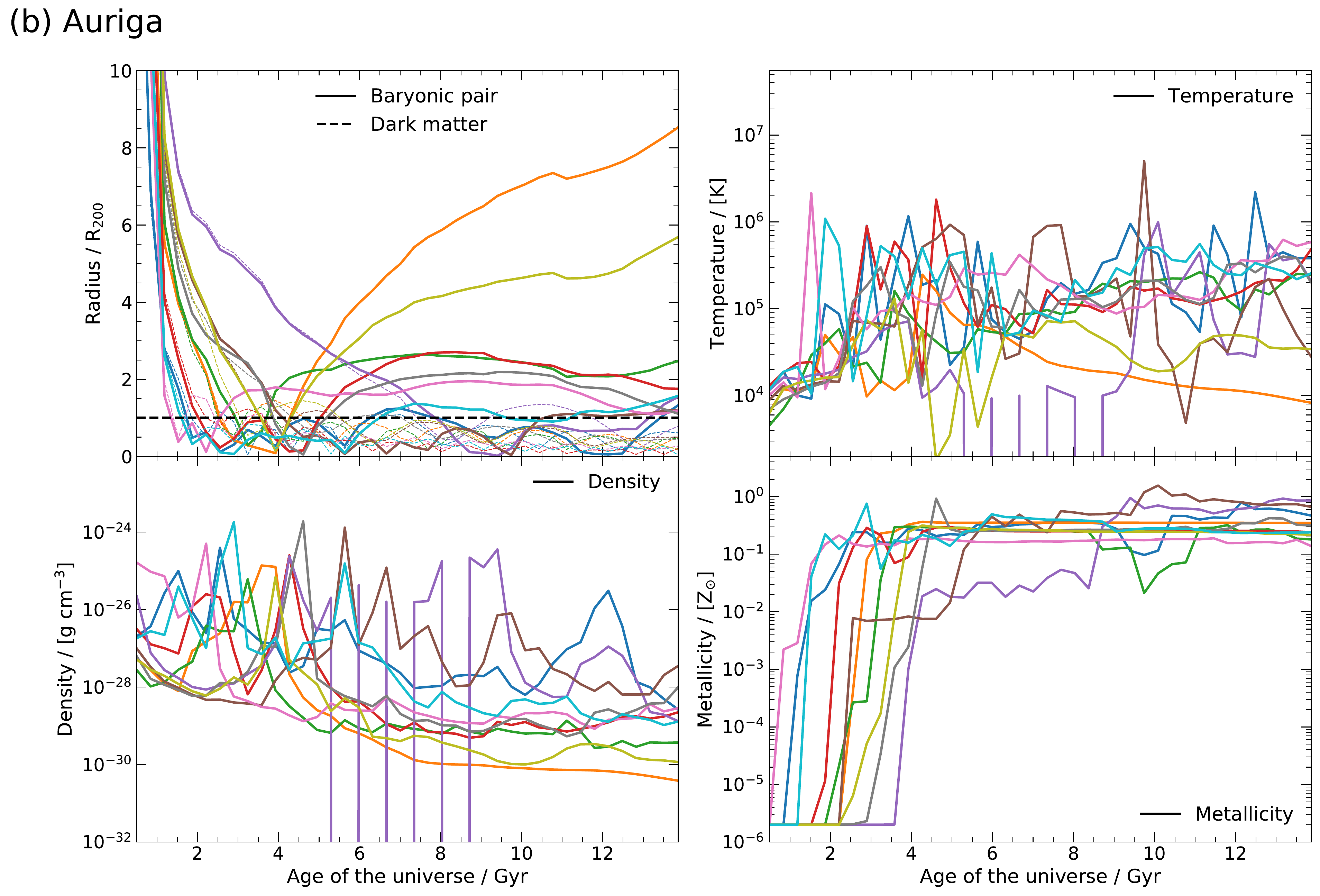}
    \caption{The evolution of 10 randomly selected halo baryonic
      tracer particles classified as `ejected' from an (a) \apostle\
      and (b) \auriga\ halo. For both (a) and (b), the top-left,
      top-right, bottom-left, and bottom-right panels show the radius
      from halo centre, mean temperature, density and metallicity as a
      function of time. The particle properties are sampled at $60$
      snapshots linearly spaced as a function of the age of the
      universe. The radius is given in units of $\rvir(z)$. The
      top-left panel also includes the radial position of the dark
      matter counterpart of each baryon particle; these are
      illustrated with a dashed line of the same colour. In the
      bottom-right panel, which shows the metallicity, there are
      several lines which lay on top of one another and thus are not
      visible. The same random particle is identified in each panel
      by the same coloured line.}
    \label{fig:ejected}
\end{figure*}

In this section we carry out a detailed analysis, on a particle-by-particle
basis, of the differences in the present-day baryon contents of the \apostle\
and \auriga\ Local Group analogues.
As we described in Section~\ref{sec:tracers_and_pmatching}, each dark matter
particle is assigned a gas particle, or a tracer particle, which shares the same
Lagrangian region in the initial conditions. In the absence of baryonic physics,
these particles would evolve purely under gravity and would thus end up in the
same halo as their dark matter counterparts. We refer to these as 'predestined'
particles. However, hydrodynamics and feedback will significantly alter their
fate.

We start our analysis by identifying all the dark matter particles present in
each of the primary haloes at $z=0$. We then identify the baryonic companions of
these dark matter halo particles and categorise their $z=0$ state, as
illustrated in Fig.~\ref{fig:missing_halo_baryons_categorised}, classifying them
into the following four categories:
\begin{itemize}
  \item \textbf{Halo gas:} gas particles within $\rvir$ of the primary halo 
  \item \textbf{Halo stars:} star particles within $\rvir$ of the primary halo 
  \item \textbf{Ejected gas:} gas particles which are outside 
    $\rvir$ at $z=0$ but had closest approach radii smaller than 
    $\rvir$ at a previous redshift 
   \item \textbf{Impeded gas:} gas particles which are outside 
     $\rvir$ at $z=0$ and had closest approach radii larger than 
     $\rvir$ at all previous redshifts. 
\end{itemize}
It is important to note that Fig.~\ref{fig:missing_halo_baryons_categorised}
does not include baryons in the present-day halo if the dark matter counterpart
is not in the halo. These baryons make a negligible contribution to the halo
baryon mass. 

There are several important differences in
Fig.~\ref{fig:missing_halo_baryons_categorised} between \apostle\ and \auriga.
About $ 35\%$ of the baryonic pairs of the $z=0$ dark matter halo particles lie
inside the primary halo of the \apostle\ simulation, whereas in the \auriga\
simulations approximately $70\%$ of the baryon counterparts are within $\rvir$.
The baryons that lie within the halo are split between stars and gas in
a roughly $1:1$ ratio in both simulations.

While the fate of the retained baryon counterparts is similar in the two
simulations, the evolution of the absent baryons is considerably different. In
\apostle\ we find that almost half of the baryon counterparts which are missing
\textit{never entered the halo}, whereas in \auriga\ almost $\approx 90\%$ of
the absent baryons entered the halo before being ejected, presumably by SNe or
AGN feedback.

\subsection{Ejected gas\label{sec:ejection}}

In Fig.~\ref{fig:ejected} (a) and (b) we show the evolution of 10 randomly
selected \predestined\ halo baryons which have been classified as `ejected'. The
particles are sampled at $60$ snapshots linearly spaced as a function of the age
of the universe, corresponding to a temporal resolution of $\approx 200
\units{Myr}$. 

In both simulations we see that the radial trajectories of both the gas and
paired dark matter particles are generally very similar at early times. As the
dark matter and baryon counterparts get closer to the main halo, their paths
deviate and there is a tendency for gas accretion to be delayed relative to dark
matter.

Following accretion into the halo, the trajectory of the gas begins to differ
significantly from that of the dark matter. This deviation is caused by
hydrodynamical forces which determine the subsequent evolution of the baryons.
Most of the baryonic particles sampled in Fig.~\ref{fig:ejected} (a) and (b)
reach a high density upon accretion and increase their metallicity. This
behaviour is consistent with halo gas that has cooled, been accreted onto the
central galaxy and later ejected. 

Focusing first on \apostle, Fig.~\ref{fig:ejected} (a), we see that most of the
randomly selected `ejected' particles were ejected over $6 \units{Gyr}$ ago. In
general, these particles reach a maximum radius, and near-constant, separation
of $\approx (3-8) \rvir$ very quickly. At the present day, this corresponds to
a physical distance of $\approx (500-1500) \units{kpc}$ and is consistent with
the conclusions of Section~\ref{sec:baryon_evolution} which reveals a baryon
deficient Local Group on scales of up to $2 \units{Mpc}$. There also appears to
be a general trend in that the earlier the gas ejection, the greater its radial
separation from the primary galaxy.

Fig.~\ref{fig:ejected} (a) shows that prior to being ejected, most of the
\apostle\ gas was cold but then underwent a sudden temperature increase, with
maximum temperatures regularly exceeding $10^{7} \units{K}$. The ejection of the
gas from the halo rapidly follows the temperature increase. This process most
likely proceeds as follows: the cold gas is part of the galaxy's ISM; some of it
is heated to a temperature $T \geq 10^{7} \units{K}$ either directly by SNe
feedback or indirectly by interaction with a SNe-heated particle. The hot gas
then gains energy and is accelerated to a velocity that exceeds the escape
velocity of the halo. This gas can then escape beyond $\rvir$ and join the
low-density IGM where it remains until the present day. It is interesting that
much of the ejected gas does not have a maximum temperature, $T = 10^{7.5}
\units{K}$. This indicates that this gas was not directly heated by SNe but
rather by interactions with SNe-heated gas.

In Fig.~\ref{fig:ejected} (b) we see that the evolution of both baryons and dark
matter is similar in \auriga\ and \apostle\ before ejection. However, in
\auriga, the ejected gas typically reaches a maximum radial distance of $\sim
3 \rvir$ before turning back and falling into the halo by the present day. Thus,
the `ejected' gas in \auriga\ likely has a shorter recycling timescale than the
gas in \apostle, much of which never reenters the halo. However, gas that was
ejected at early times and re-accreted is not, by definition, included in
Fig.~\ref{fig:ejected} (b). Short recycling times in \auriga\ were first
reported by \cite{grand:2019} who show the model gives rise to efficient
galactic fountains within the inner $\sim 30 \units{kpc}$, with median recycling
times of $\sim 500 \units{Myr}$ in MW-mass haloes.

\begin{figure*}
    \centering
    \includegraphics[width=0.85\textwidth]{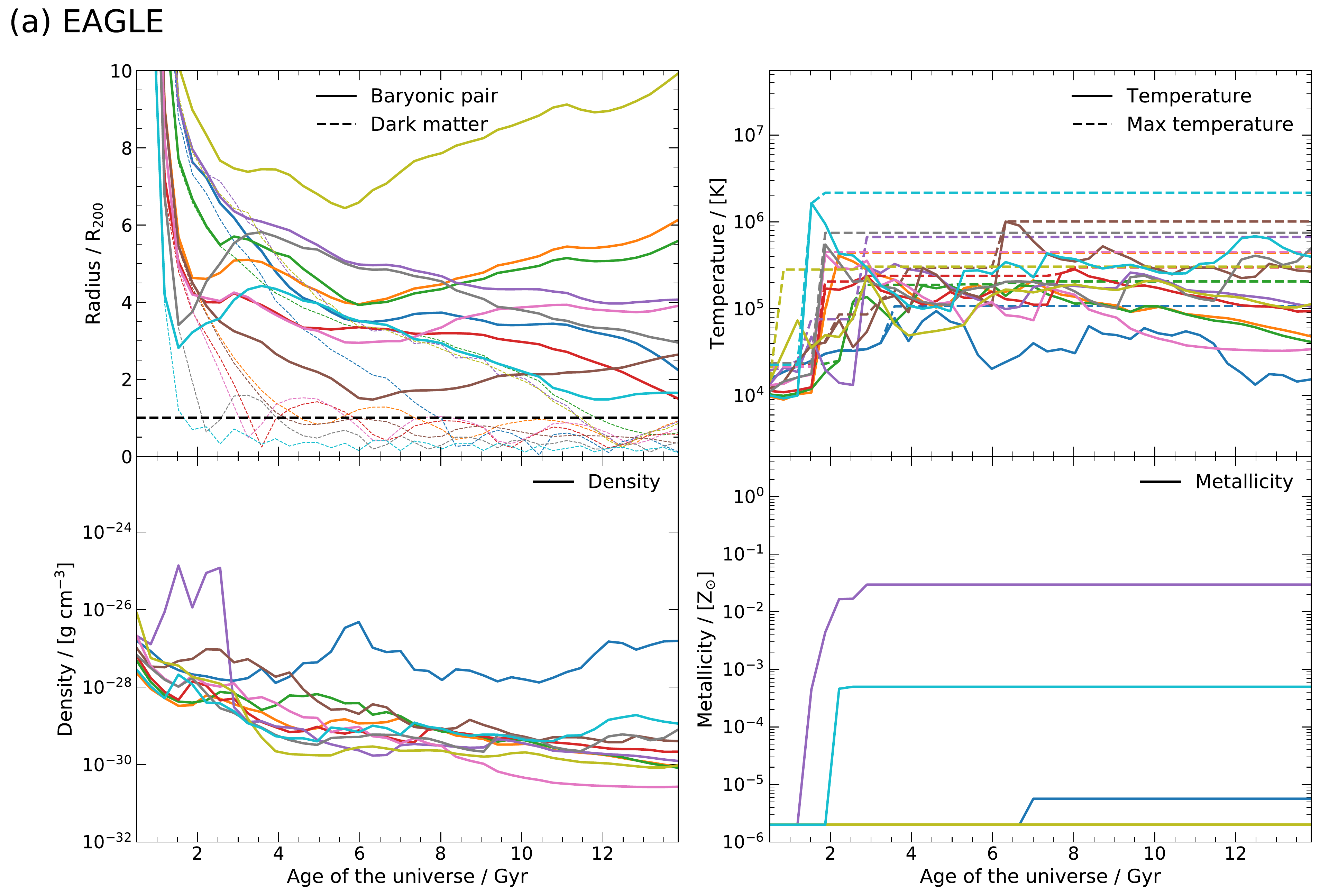}
    \includegraphics[width=0.85\textwidth]{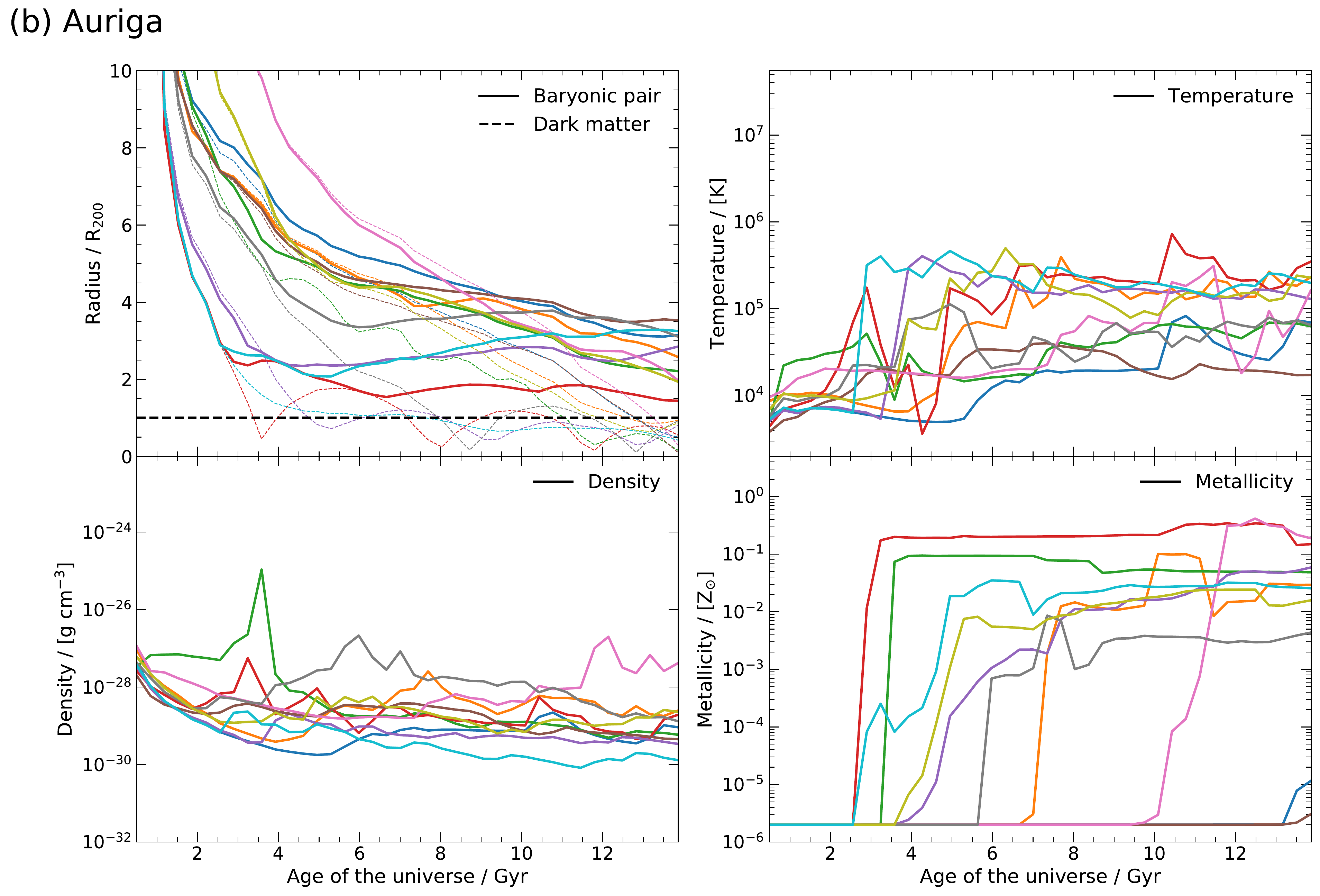}
    \caption{The evolution of 10 randomly selected halo baryonic tracer
        particles classified as `impeded' from an (a) \apostle\ and
        (b) \auriga\ halo. For both (a) and (b), the top-left, top-right,
        bottom-left, and bottom-right panels show the radius from 
        halo centre, temperature, density and metallicity as a
        function of time. The particle properties are sampled at 
        $60$ snapshots linearly spaced as a function of the age of the
        universe. The radius is normalised to the (time-dependent) 
        $\rvir$. The top-left panel also includes the radial position
        of the dark matter counterpart of each baryon particle; 
        these are illustrated with a dashed line of the same
        colour. In the bottom-right panel, which shows the
        metallicity, there are several lines which lay on top of one
        another and thus are not visible. As with Fig.~\ref{fig:ejected} the
        same random particle is identified in each panel by the same coloured line.
    }
    \label{fig:impeded}
\end{figure*}

\subsection{Impeded gas}

We now analyse the evolution of ten randomly selected \predestined, baryonic
particles which have \textit{never} entered the primary halo. These are plotted
in Fig.~\ref{fig:impeded} (a) and (b) for \apostle and \auriga, respectively.
In Fig.~\ref{fig:impeded}(a) we see that about half the randomly-selected
particles in \apostle make their initial approach at redshift, $z \sim 2-3$. As
the dark matter particles are accreted into the halo, their baryon counterparts
start being impeded at a radius $\sim 4 \rvir$. The subsequent fate of these
particles can differ substantially. About half of them remain at a distance
$\geq 3 \rvir$ until the present day, whereas the other half continue to
approach the primary halo and are almost accreted by $z=0$.

The (maximum) temperature of the \predestined\ impeded particles in \apostle\ is
shown in the top-right panel of Fig.~\ref{fig:impeded} (a). The overall
temperature evolution of all impeded particles is quantitatively similar.
Initially, the gas is at low temperature, $\approx 10^{4} \units{K}$, but as it
approaches the halo, it is subject to an almost instantaneous temperature rise
to $\approx 10^{6} \units{K}$. Since the maximum temperature of the gas is
always well below $10^{7.5} \units{K}$, this rise is not the result of direct
SNe or AGN heating. Furthermore, before closest approach, these gas particles
also have low density and metallicity, with one or two exceptions. These
properties are consistent with pristine gas within the IGM, thus confirming that
direct feedback did not heat these particles.

In Fig.~\ref{fig:impeded} (b) we see that the accretion time of the dark matter
counterparts of the ten randomly selected impeded baryons in \auriga\ is recent,
within the last $\approx 4 \units{Gyr}$, for most particles. In \apostle we saw
that before closest approach, the baryon particles were relatively unenriched,
cold and at low density showing no sign of interaction with nearby galaxies.
However, as seen in the bottom right panel of Fig.~\ref{fig:impeded} (b), much
of the impeded gas in \auriga\ is enriched; this suggests the cause of the
impediment could be local interactions.

\begin{figure}
    \centering
    \includegraphics[width=1.00\linewidth]{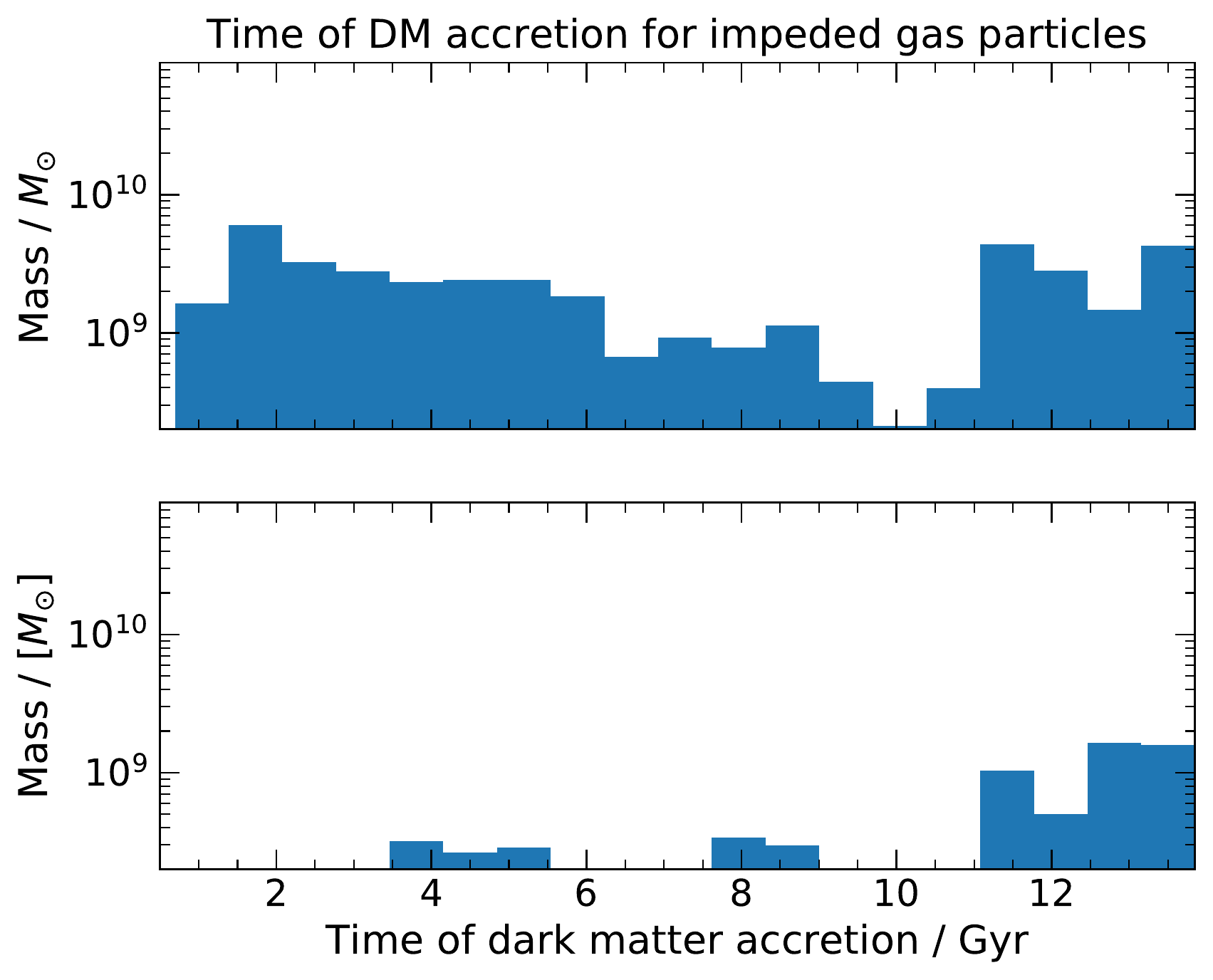}
    \caption{Mass-weighted histogram of the age of the universe at the
      time of accretion of the dark matter particles associated with
      \predestined\ baryon particles that have been
      \textit{impeded}. The top panel shows results for the
      AP-S$5$-N$1$ halo in \apostle\ and the bottom panel for
      \auriga. This shows that cosmic gas accretion is significantly
      impeded at all times in \apostle, whereas \auriga\ only impedes
      cosmic gas accretion, relative to dark matter accretion, at late
      times.
    }
    \label{fig:dm_accretion_time_for_impeded}
\end{figure}

In Fig.~\ref{fig:dm_accretion_time_for_impeded} we show the mass-weighted
distribution of first accretion times of all dark matter particles associated
with \predestined\ baryon particles classified as `impeded'. The \apostle\
simulations have a weakly bimodal distribution: the dark matter associated with
impeded gas was accreted either at early or at late times. In the \auriga\
simulations the late accreted population dominates.

The presence of hot quasi-hydrostatic haloes can impede gas accretion at late
times. haloes of mass $\gtrapprox 10^{12} \units{\msun}$ have massive hot gaseous
coronae. These gaseous haloes, which often extend beyond $\rvir$, can exert
pressure on the accreting gas and thus delay its accretion. The accretion of the
collisionless dark matter is, of course, not impeded.
The process impeding gas accretion at early times and late time in \apostle\ is
likely to be similar. However, the progenitors of the present-day primary haloes
are not massive enough at high redshift to support massive atmospheres of
primordial gas to explain the observed scale of suppressed accretion so most of
the gaseous atmosphere must be gas that was reheated and ejected.  

\begin{figure*}
    \centering
    \includegraphics[width=1.0\textwidth]{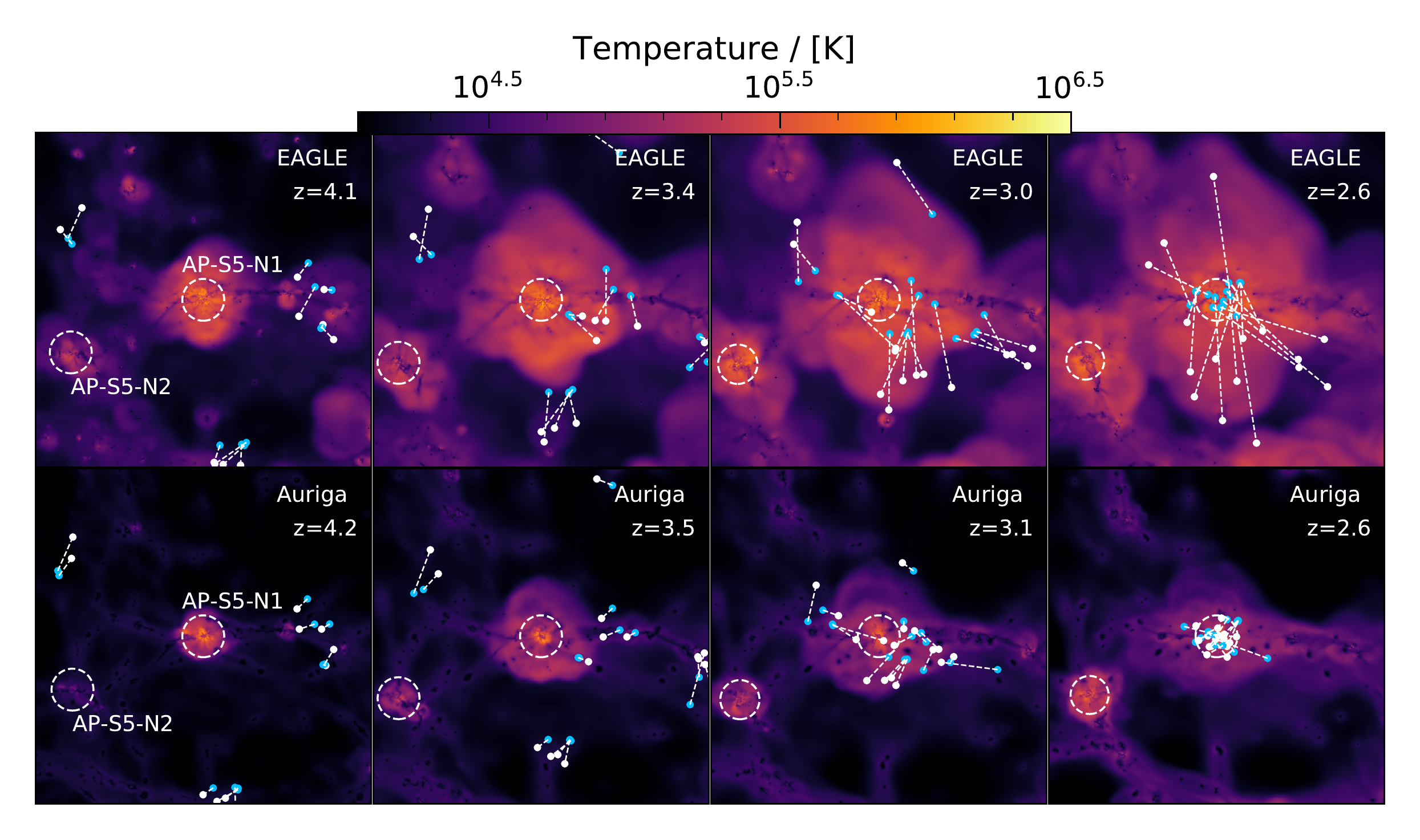}
    \caption{Mass-weighted temperature projections through a region of
      size $(16 \times 16 \times 24)~ \rvir$ centred on the location
      of the progenitor of the primary galaxy in AP-S$5$-N$1$ at four
      redshifts, $z\sim4, 3.5, 3, 2.5$.  Results are shown for
      both \apostle\ (top row) and \auriga\ (bottom row).  The white
      and blue circles show the positions of sixteen randomly selected
      baryon-dark matter counterparts which are `impeded' at  early
      times in \apostle. The same counterparts, originating from the
      same Lagrangian region, are shown in \auriga. The white dotted
      lines connect the baryon/dark matter counterparts. The white dashed
      circles identifies the $\rvir$ of the two main haloes.}
    \label{fig:temp_tracers}
\end{figure*}

The explanation is provided in Fig.~\ref{fig:temp_tracers} which shows
temperature projections of the AP-S$5$-N$1$ halo at four times between $z=3.5$
and $z=2$ in both \apostle\ and \auriga. We overlay the positions of sixteen
randomly selected baryon/dark matter pairs chosen so that they were `impeded' at
early times in \apostle. The same pairs, originating from the same Lagrangian
region in \auriga, are also overlaid. The blue circles show the $x, y$ positions
of the dark matter and the white circles those of the baryons. The white dotted
line connects the baryon/dark matter pairs to help visualise the differences in
the evolution of the two species.

We see that in \apostle\ the MW progenitors are encased in a hot, $\approx
10^{6} \units{K}$, corona of gas that occupies a volume of radius twice as large
as that of the \auriga\ counterparts. This gas halo consists of a mixture of
accreting primordial gas and hot outflowing winds fuelled by feedback from the
central galaxy.  As shown in Fig.~\ref{fig:ejected}, gas particles in the ISM
can be heated by SNe feedback, generating a hot, outflowing wind that can reach
distances of up to $4 \rvir$ in less than $1 \units{Gyr}$. This outflowing gas
interacts with the accreting gas, applying an outward force sufficient to delay
or prevent accretion. This leads to large amounts of \predestined\ gas `impeded'
from accreting at early times in \apostle\ (as seen in
Fig.~\ref{fig:missing_halo_baryons_categorised}). The overlaid particles in
Fig.~\ref{fig:temp_tracers} succinctly demonstrate this process.

In contrast, the \auriga\ haloes have hot gaseous components that barely extend
beyond $\rvir$ and do not evolve in time significantly. As the hot component in
\auriga\ is less massive and cooler than in \apostle, the dark matter/baryon
pairs evolve similarly until they reach $\sim 1.5 \rvir$. At radii $\leq 1.5
\rvir$ gas accretion is delayed relative to the dark matter by hydrodynamical
forces. However, it appears less than $5\%$ of gas is completely prevented from
accretion.

\subsection{The fate of impeded and ejected baryons}

\begin{figure*}
    \centering
    \includegraphics[width=1.00\textwidth]{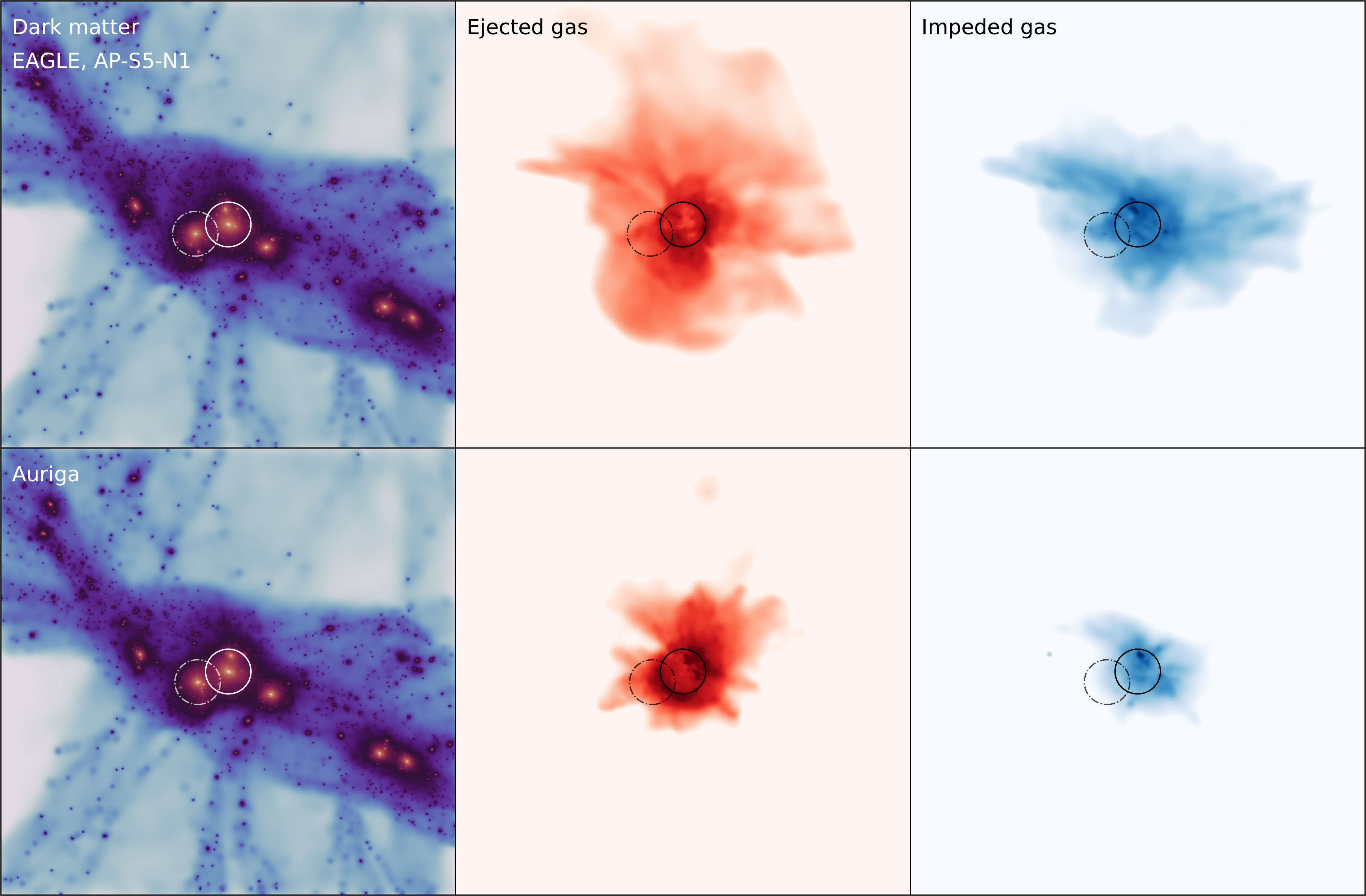}
    \caption{The $z=0$ projected density of dark matter (left),
      ejected baryons (center) and impeded baryons (right) for the
      AP-S$5$-N$1$ halo in both \apostle\ (top) and \auriga\
      (bottom). The projected cuboid has dimensions of
      $20 \times 20 \times 8 \units{\rvir}$ in the $x$, $y$ and $z$
      directions, respectively. The solid white/black
      circles indicate $\rvir$ of the halo AP-S$5$-N$1$ and the
      dashed-dotted circles that of AP-S$5$-N$2$. }
    \label{fig:missing_halo_baryons_projected}
\end{figure*}

Fig.~\ref{fig:missing_halo_baryons_projected} shows the projected dark matter
and gas density of both `ejected' and `impeded' baryons that were \predestined\
to end up in the primary halo but are missing.
In the \apostle\ simulations, the present-day distribution of these missing
baryons is different for the `impeded' and `ejected' components. The impeded gas
is elongated roughly along the $x$-axis which, as can be seen in the projected
dark matter distribution, traces a local filament. By contrast, the ejected gas
tends to be elongated along the $y$-axis, that is, perpendicular to the
filament.

The impeded gas flows to the halo along the filament and would have been
accreted by the halo had the pressure of the hot halo not impeded it. Thus, it
remains in the filament, centred around the halo. The ejected material, on the
other hand, finds the path of least resistance, which is perpendicular to the
filament: along the filament direction the wind encounters relatively high
density gas while in the perpendicular direction, the density and pressure of
the surrounding medium drop rapidly. As a result, gas ejected perpendicular to
the filament can reach larger radii, giving rise to the elongated distribution
seen in Fig.~\ref{fig:missing_halo_baryons_projected}.

An interesting detail in Fig.~\ref{fig:missing_halo_baryons_projected} is the
transfer of baryons between the M31 and MW analogues. This is due primarily to
the effect of gas ejection from both galaxies which can cause
cross-contamination. This process is an example of halo gas transfer,
\citep[see][]{borrow:2020} and indicates that the proximity of M31 may have
influenced the evolution of the MW, although the amounts of gas transferred are
very small.

While the two simulations predict very different morphologies for the ejected
baryons, these differences are likely undetectable in the real universe because
of the low density of ejected material.

\section{Understanding the differences}\label{sec:subgrid_diff}

\begin{figure*}
    \centering
    \includegraphics[width=0.95\textwidth]{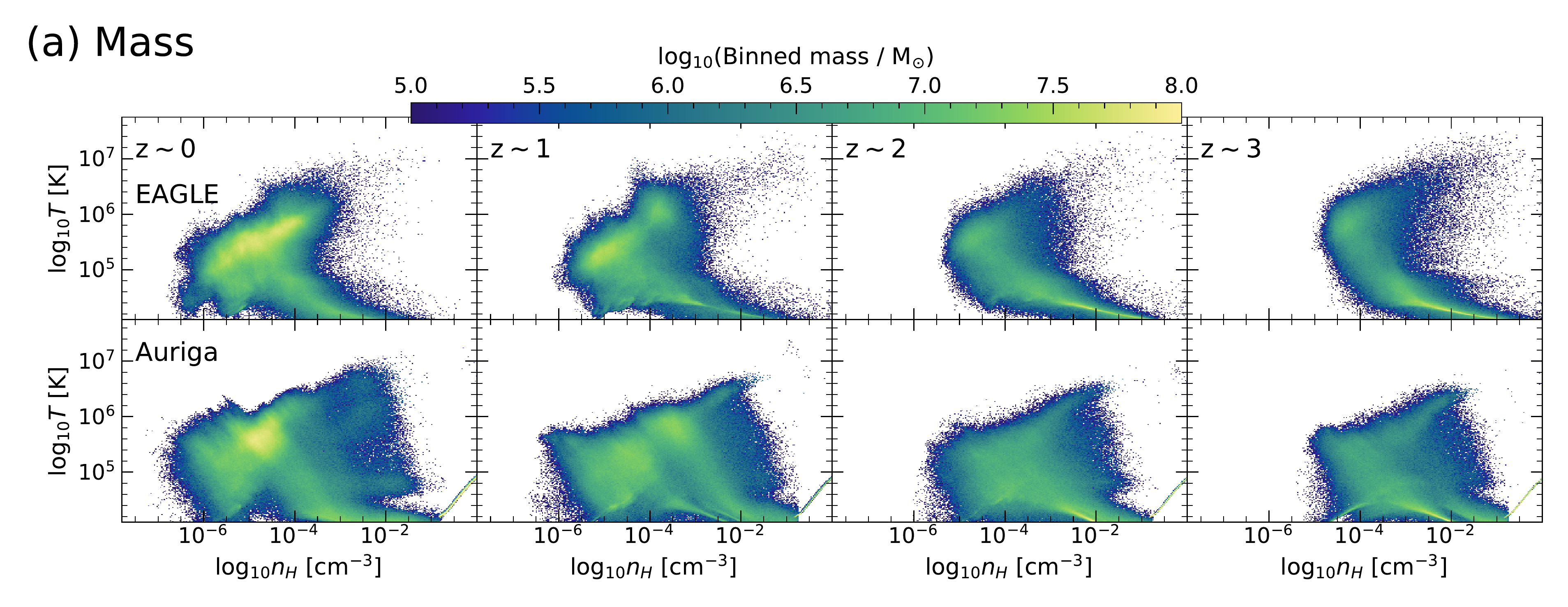}
    \includegraphics[width=0.95\textwidth]{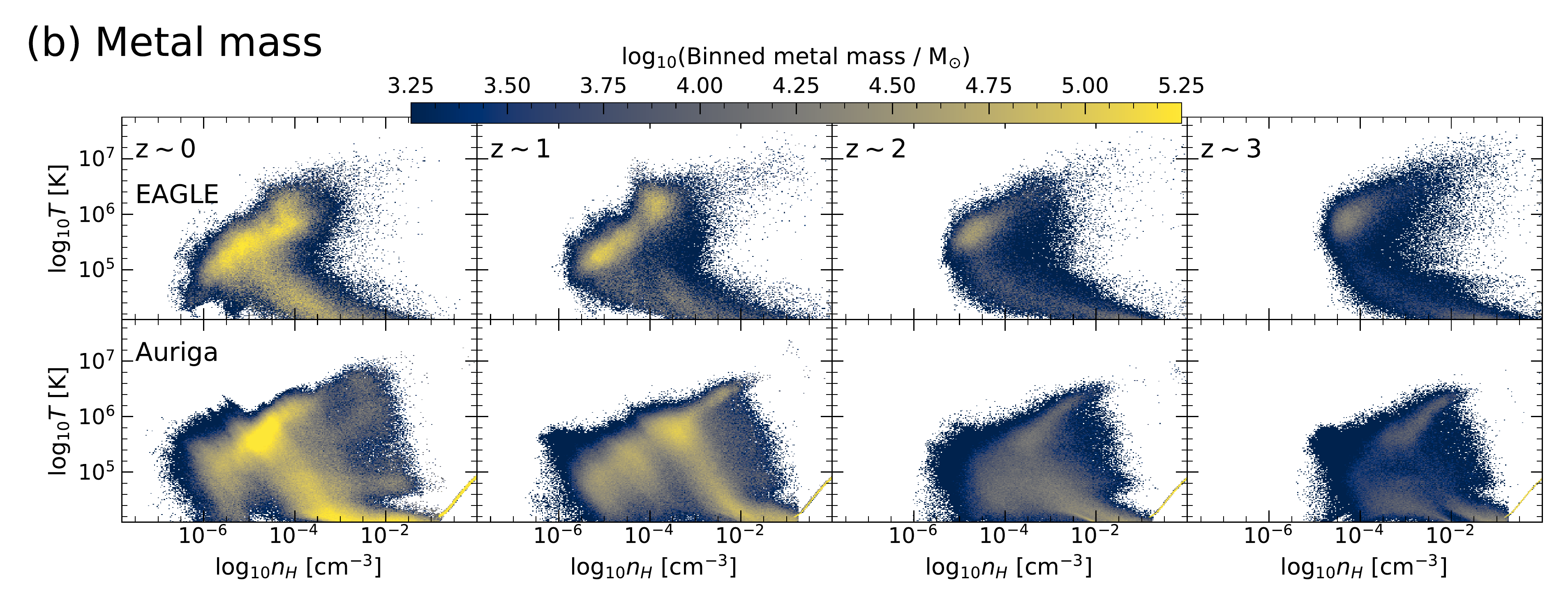}
    \includegraphics[width=0.95\textwidth]{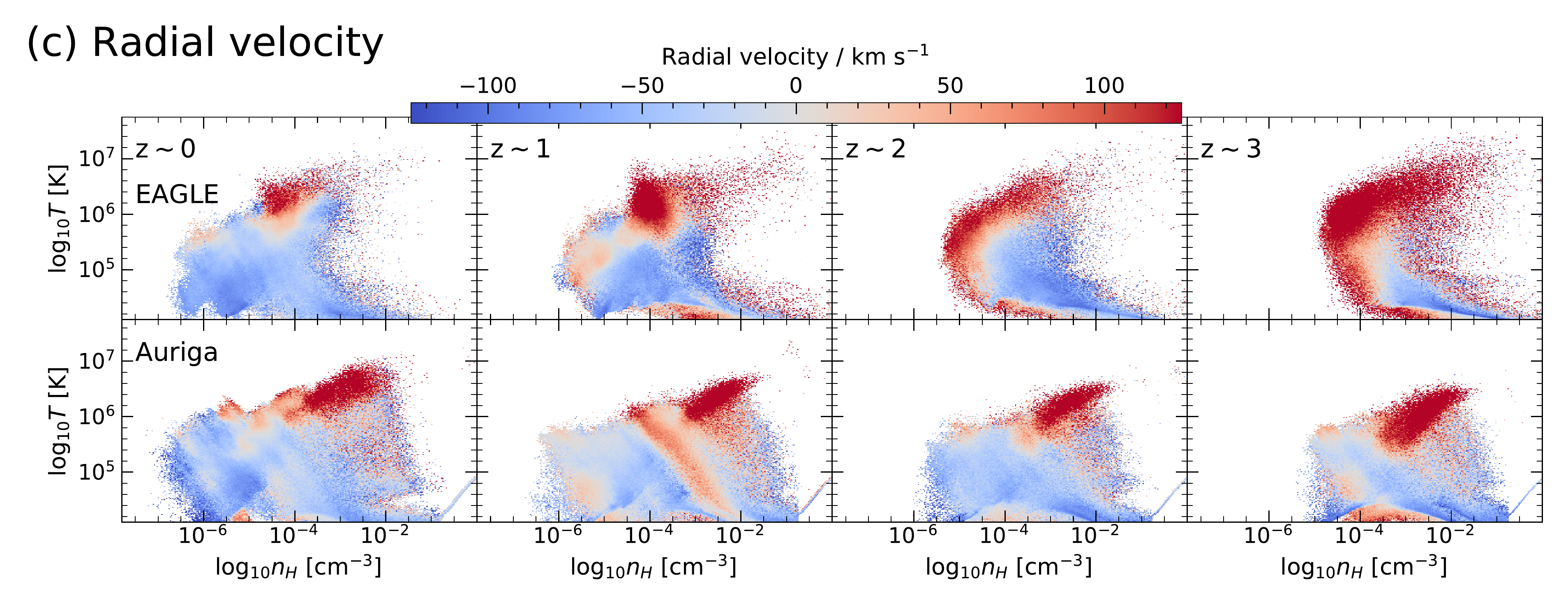}
    \caption{Gas density-temperature diagrams for a primary halo,
      AP-S$5$-N$1$, in both \apostle\ (top row) and \auriga\ (bottom
      row). The histograms are coloured according to (a) mass, (b)
      metal mass and (c) mass-weighted radial velocity. They include
      all gas with temperature $T > 10^{4} \units{K}$ that is within
      radius, $r < 3 \rvir$, of the primary halo. The four columns
      show the gas distributions at four redshifts,
      $z \sim 0,~1,~2, ~3$, from left to right, respectively.  The
      histograms are generated with $300$ logarithmically spaced bins in the
      density range, $10^{-8} - 1 \units{cm^{-3}}$, and temperature
      range, $10^{3.5} - 10^{7.5} \units{K}$.}
    \label{fig:density_temperature_phase}
\end{figure*}

In Fig.~\ref{fig:density_temperature_phase} we show histograms of the gas
density and temperature weighted by mass, metal mass and radial velocity for
both \apostle\ and \auriga\ at four redshifts. All gas particles/cells within
a sphere of radius $3~\rvir$ around the centre of the primary halo of
AP-S$5$-N$1$ are included. As in Section~\ref{sec:baryon_evolution}, we focus on
this particular example but our general results are valid for all haloes in our
sample.

We can see in the figure the gas responsible for impeding accretion in \apostle\
at $z=2-3$: it is hot, $\geq 10^{6} \units{K}$, low density, $\sim 10^{-4}
\units{cm^{-3}}$, slightly metal-enriched and outflowing with a mean radial
velocity exceeding $100 \units{km~s^{-1}}$ relative to the centre of mass of the
halo. This gas component is visible from $z=3$ until $z=1$. In \auriga\ gas with
similar temperature and density is less enriched and is not outflowing. There is
some hot enriched gas with large outflow velocities at all redshift, but this
gas appears to cool and mix with inflowing material as there is no evidence of
less dense and slightly cooler outflows.

At higher redshift, the majority of the metals in
Fig.~\ref{fig:density_temperature_phase}(b) reside in hot, diffuse gas. As shown
by Fig. \ref{fig:density_temperature_phase} (c) this material is outflowing.
Interestingly, we do not see any evidence of a significant cooler metal
component developing until about $z=1-2$. When this cooler metal component
appears, it is radially inflowing, suggesting the recycling of earlier outflows.

In \auriga\ we first see a population of metal-enriched gas that is both hot and
dense at approximately $z=3$. By $z=2$, this enriched gas appears to have cooled
and increased in density. This is inferred from distribution of metals. These
features suggest the presence of galactic fountains even at high-redshift. The
\auriga\ haloes also contain a component of very dense gas, $n_{\rm H} \geq
10^{-2} \units{cm^{-3}}$, with temperature  ranging between $T \approx 10^{4}
- 10^{7} \units{K}$.  This is a further indication of efficient galactic
fountains: the high densities lead to short cooling times, of order $\sim 200
\units{Myr}$, and, for this substantial amount of gas to be present, it must be
continuously replenished by the heating of dense gas by feedback. 

The differences in the nature of outflows in \apostle\ and \auriga\ could be due
to differences in the SNe subgrid models. The \apostle\ SNe feedback model
specifies the temperature increase of gas particles. In this model, SNe energy
is effectively saved up and released in concentrated form stochastically. This
technique means that gas is heated to higher temperatures less frequently, thus
preventing the over-cooling problem \citep{vecchia:2012}.
Whereas in \auriga, the energy injected per unit mass of SNII is fixed. This
difference in the model causes gas in \apostle\ to reach higher temperatures
post-SNe feedback.

The total cooling efficiency of moderately enriched gas has a local minimum
around a temperature of $10^{7} \units{K}$ and increases quite steeply both with
increasing and decreasing temperature \citep[see e.g. Fig. 9 of ][]{baugh:2006}.
Thus, a post-feedback temperature lower than $10^{7} \units{K}$ leads to both
a higher cooling rate and lower thermal energy. When these effects are combined,
the cooling time can be reduced by an order of magnitude or more. The lower
post-feedback temperature in \auriga\ could allow SNe-heated gas to radiate
a significant fraction of the injected energy on a timescale of order several
hundred million years. This reduced cooling time would facilitate short
recycling times for gas in \auriga\ and prevent the build up of a hot,
SNe-fueled atmosphere at high redshift.

In \apostle, by contrast, feedback heats the gas to $10^{7.5} \units{K}$, and
thus radiative cooling is relatively inefficient. Figure~8 of \cite{kelly:2020}
shows the density and temperature of gas particles $500 \units{Myr}$ before and
after SNe heating in the \apostle\ reference simulation. That figure shows that
gas particles that experience SNe feedback decrease their density by over two
orders of magnitude within $\leq 20 \units{Myr}$ of being heated. This expansion
prevents efficient radiative cooling and also makes the gas buoyant so that it
is accelerated out through the halo \citep{bower:2018}.

\begin{figure*}
    \centering
    \includegraphics[width=0.95\textwidth]{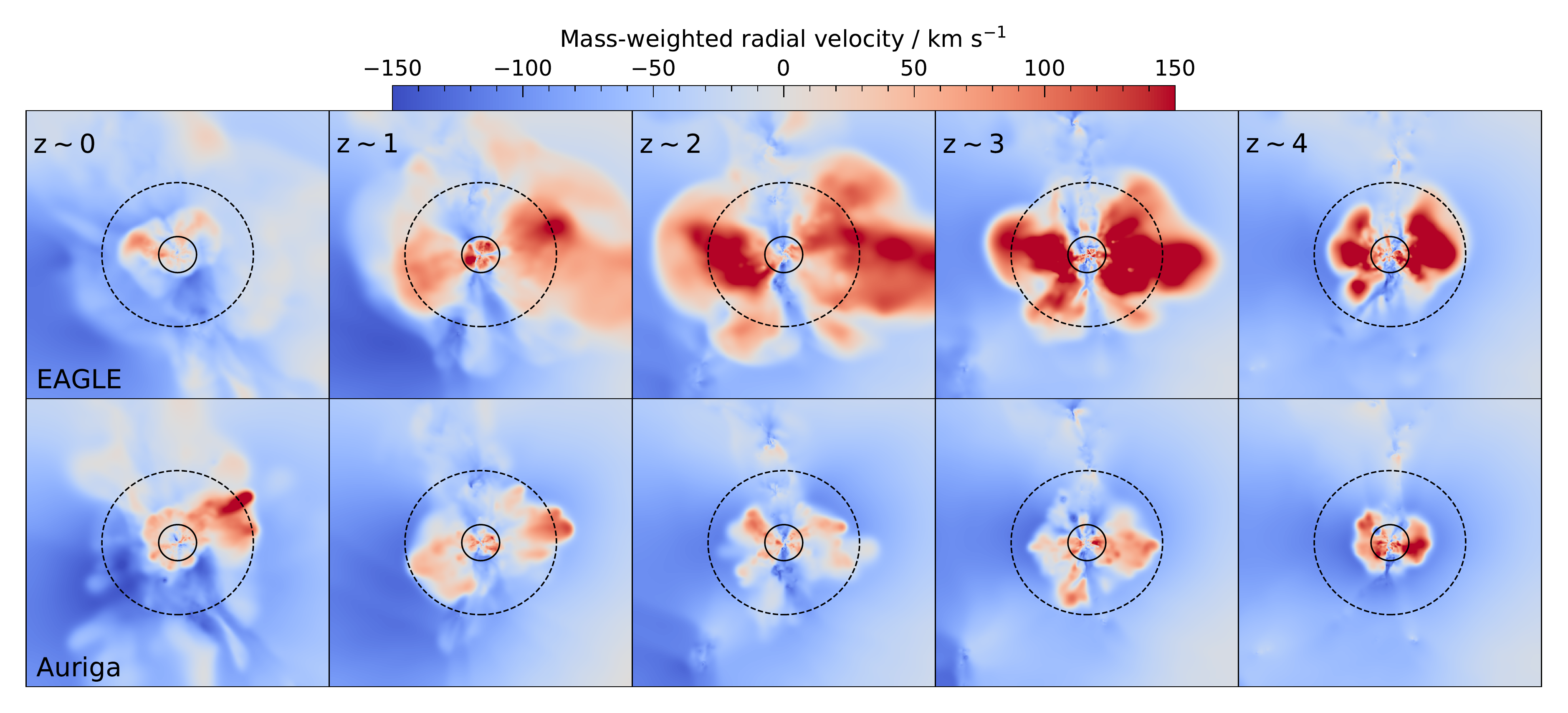}
    \caption{Mass-weighted projection of the radial gas velocity for
      AP-S$5$-N$1$ in both \apostle\ (top row) and \auriga\ (bottom
      row) at five redshifts, $z \sim 0,~1,~2,~3,~4$, from left to
      right, respectively. The region of these projections has size
      $(16 \times 16 \times 2)~\rvir$ in the $x$, $y$ and $z$
      directions, respectively, where the $z$-axis is into the
      page. The solid black circle indicates $\rvir$ and the dashed
      back curve $4 \times \rvir$.  }
    \label{fig:velocity_slice}
\end{figure*}

Fig.~\ref{fig:velocity_slice} shows a mass-weighted projection of the radial
velocity of the gas in AP-S$5$-N$1$ for both \apostle\ (top row) and \auriga\
(bottom row) at five redshifts, $z \sim 0,~1,~2,~3,~4$. It is clear from these
projections that rapid outflowing material extends well beyond the halo,
reaching distances of $\sim 4 \rvir$ at $z=4$ in \apostle. These strong,
halo-wide outflows are readily visible until $z=1$. By contrast, in \auriga\,
the mass-weighted radial velocity of the outflows is much slower, and their
spatial extent is smaller. These radial velocity projections indicate that
\apostle\ haloes experience strong outflows with high covering fractions. A high
outflow velocity, combined with a large covering fraction, can impede
cosmological gas accretion on the scale of the entire halo. By contrast, in
\auriga\, we see less extended outflows because a significant fraction of the
SNe-heated material cools and is efficiently recycled near the centre of the
halo.
\cite{voort:2021} shows that the magnetic fields, included in the \auriga\
simulations, can reduce the outflow velocities of gas around the central
galaxies. This extra pressure from magnetic fields in \auriga\ could be
contributing to the reduced gas ejection from the halo compared to that found in
\apostle.

\begin{figure}
    \centering
    \includegraphics[width=0.95\linewidth]{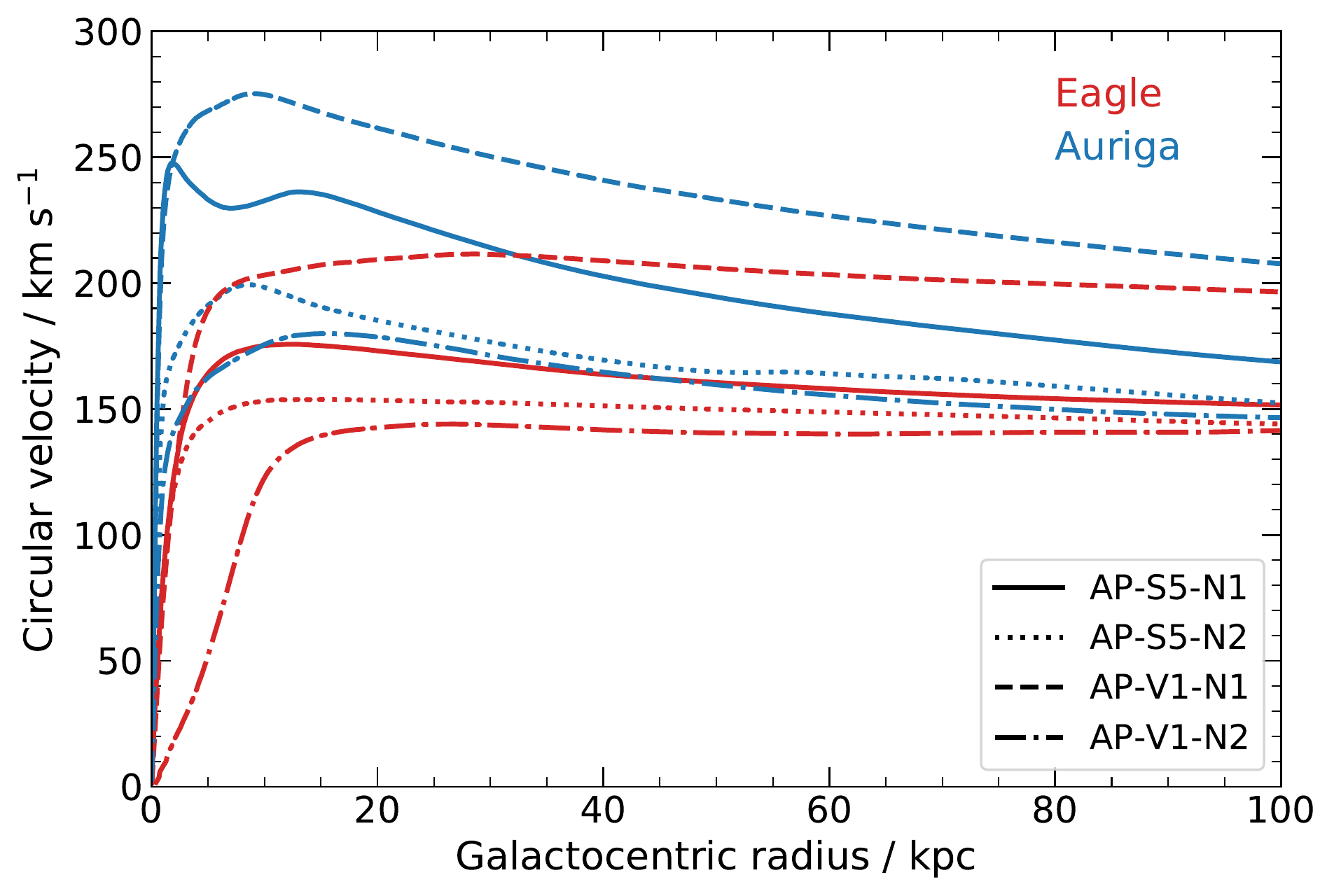}
    \caption{The approximate circular velocity of the four primary haloes in both
        \apostle\ and \auriga\ at $z=0$. The circular velocity, $V_{\rm circ}$, at
        radius $r$ is calculated from the total enclosed mass, $M(< r)$, assuming
        spherical symmetry $V_{\rm circ}^2 = G M(<r) / r$ where $G$ is the
        gravitational constant. The \apostle\ and \auriga\ simulations are shown in
        red and blue, respectively.
    }
    \label{fig:vcirc_profile}
\end{figure}

Another factor that can vary the velocity and spatial extent of outflows is the
potential depth. In Fig. \ref{fig:vcirc_profile} we show the approximate
circular velocity profiles as a function of radius in the four primary haloes in
both \apostle\ and \auriga\ at $z=0$. These approximate circular velocity
profiles are calculated from the total enclosed mass assuming spherical
symmetry.
We see that in the inner $50 \units{kpc}$ the circular velocity is
systematically higher in \auriga, by up to as much as $\sim 100 \units{km
s}^{-1}$. This difference is because the \apostle\ model is much more efficient
at driving winds from the central galaxy, propagating out to scales exceeding
the virial radius \citep{mitchell:2020a}. This process reduces the density in
the inner region, thus reducing the escape velocity and making future gas
ejection more efficient. Whereas in \auriga\ the opposite happens, as wind is
inefficient at removing gas from the central region, the central density
increases, thus deepening the potential well.

As described in Section~\ref{sec:the_sample}, the gas dynamics in \auriga\ are
followed with a moving-mesh technique. Several studies have investigated the
differences between moving-mesh and particle-based hydrodynamics techniques in
the context of galaxy formation simulations \citep{sijacki:2012, keres:2012,
vogelsberger:2012, torrey:2012, bird:2013, nelson:2013}. A general result from
these studies is that gas in moving-mesh simulations cools more efficiently than
in their particle-based counterparts. The cooling efficiency of hot gas is
artificially suppressed in particle-based simulations by spurious viscous
heating and the viscous damping of SPH noise on small scales. Furthermore,
moving-mesh simulations model energy dissipation more realistically by allowing
cascading to smaller spatial scales and higher densities \citep{nelson:2013}. As
we mentioned earlier, we expect these difference to be subdominant to the large
differences in the subgrid models between \apostle\ and \auriga.

\section{Observational tests}\label{sec:observables}

We have shown in Sections~\ref{sec:baryon_evolution} and
\ref{sec:missing_halo_baryons} that the \apostle\ and \auriga\ simulations
predict very different baryon cycles around our MW and M31 analogues.
We now turn to the observable signatures of strong outflows in \apostle\ and
galactic fountains in \auriga. 
We present mock observations of absorber column densities and dispersion measure
around our MW and M31 analogues. We aim to construct mappings between the
physical state of the baryons and real observables and, in particular, to
identify observables that may be sensitive to the differences in the gas
properties seen in the \apostle\ and \auriga\ simulations.

\subsection{Column densities}\label{sec:ions_obs}

\begin{figure}
    \centering
    \includegraphics[width=1.00\linewidth]{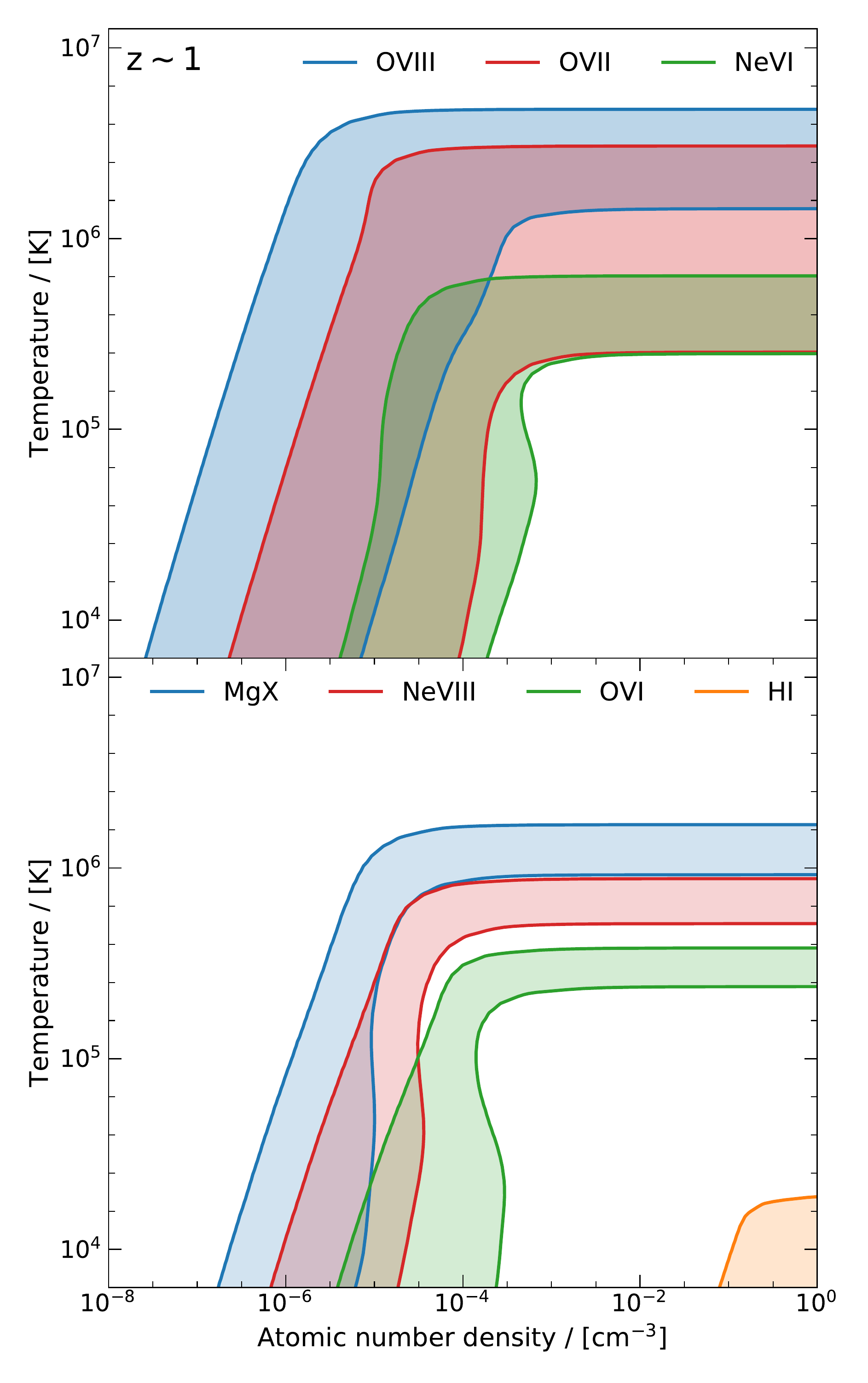}
    \caption{Temperatures and densities at which different metal
      species occur at $z=1$. The ion fractions are calculated from
      the lookup tables of \protect\cite{hummels:2017}, assuming only
      radiation from the metagalactic UV background according to the
      model of \protect\cite{haardt:2012}. These lookup tables are
      computed under the assumption of collisional ionisation
      equilibrium (CIE). The contours for each of the indicated
      species are at 10 per cent of the maximum ion fraction. The top
      panel shows \NeVI\ (green), \OVII\ (red) and \OVIII\ (blue) and
      the bottom panel \HI\ (orange), \OVI\ (green), \NeVIII\ (red)
      and \MgX\ (blue). }
    \label{fig:contour_density}
\end{figure}

We show, in Fig.~\ref{fig:contour_density}, the regions of the
density-temperature plane where different species are relatively abundant. The
contours enclose regions where each species contributes $10$ per cent of the
maximum ion fraction of the respective element in collisional ionisation
equilibrium at $z=1$. We show these contours as they highlight regions of the
density-temperature phase space where each species is likely to be detected.
This analysis is similar to \cite{wijers:2020} which analysed lower resolution,
large volume cosmological simulations using the \apostle\ galaxy formation model.

The species that probe the hottest gas, at $T \geq 10^6 \units{K}$, are
typically \OVII\, \OVIII\ and \MgX. However, there is a degeneracy as these
ionisation states are also plentiful in lower temperature, lower density gas. In
the absence of other observables, it is difficult to distinguish whether
detection of these ions is from a low density or a high density region. However,
given that, in practice, the detection of these lines requires a moderately high
column density, it is likely that any detection will be from gas at high density
and, thus, high temperature.  We also see in the figure that \NeVI, \NeVIII\ and
\OVI\ probe a cooler component, at $T \leq 10^{6} \units{K}$. This transition
temperature, $T = 10^{6} \units{K}$, is a significant threshold as it has the
potential to distinguish hydrostatic gas at the virial temperature of MW-mass
haloes from hotter gas heated by feedback energy.

\begin{figure*}
    \centering
    \includegraphics[width=0.90\textwidth]{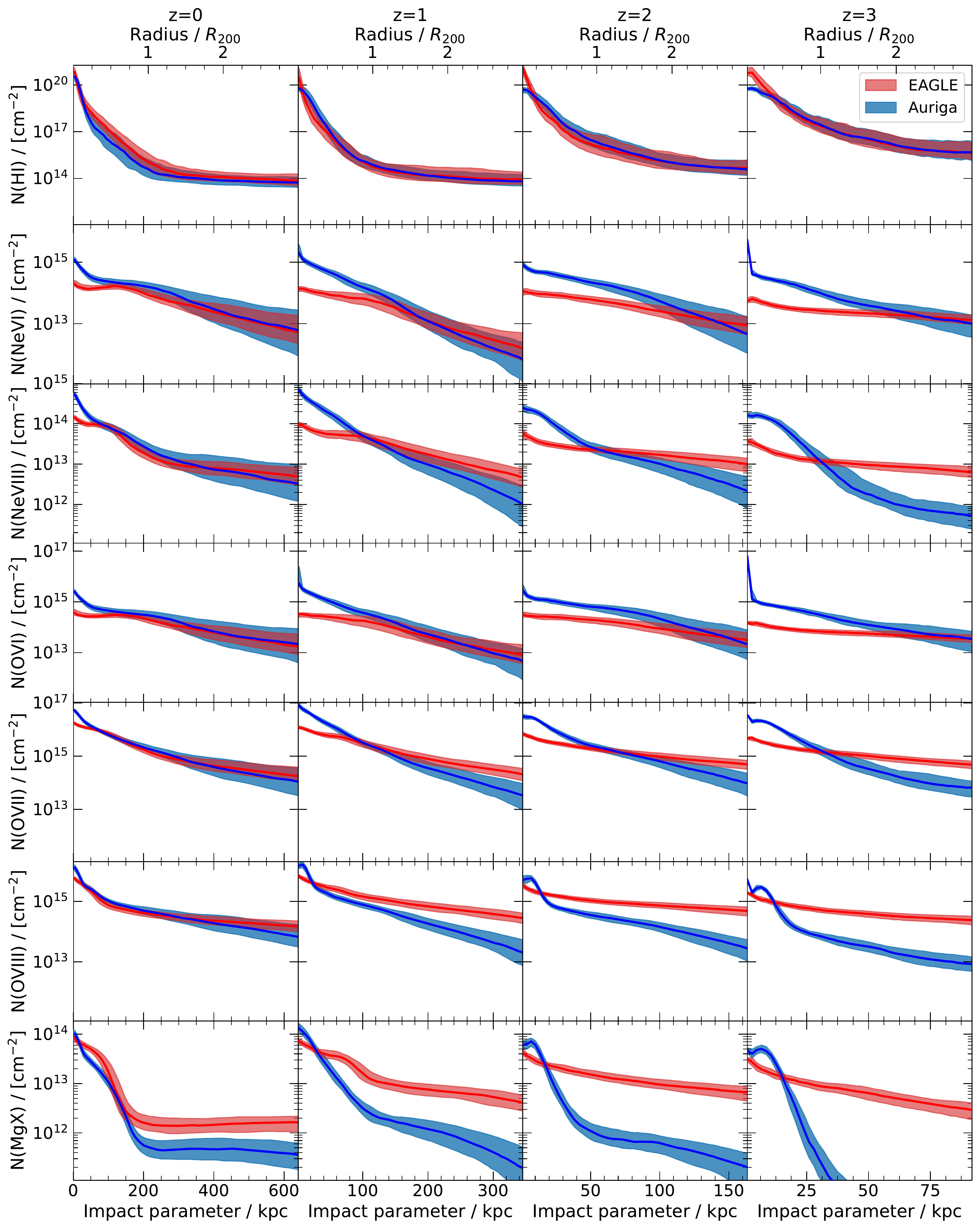}
    \caption{Atomic column number densities of several species as a
      function of impact parameter from the centre of the primary halo
      in both the \apostle\ (red) and \auriga\ (blue) simulations, out
      to $3 \rvir$ at four redshifts $z \sim 0,~1,~2,~3$ from left to
      right, respectively.  From top to bottom, the rows show results
      for \HI, \NeVI, \NeVIII, \OVI, \OVII, \OVIII\ and \MgX,
      respectively.  The column density at a fixed radius, $r$, for
      each halo is calculated by taking the median column density of
      many sightlines through a small annulus. We also compute the
      lower and upper quartiles of the column density in each
      annulus. The solid lines are the mean of the median column
      density, at each radius, for all four haloes. The shaded regions
      illustrate the range between the mean of the lower and upper
      quartiles for all of the haloes. We choose to use the median and
      quartiles as these are more comparable to a single random
      line-of-sight through a halo in the real universe.}
    \label{fig:column_density_radius}
\end{figure*}

The species \OVII\ and \OVIII\ are well suited for identifying very hot
outflows. Unfortunately, these ions are challenging to observe in the real
universe because the wavelengths of their lines are so short, $\sim 20
\units{\angstrom}$. These highly ionized stages of oxygen, the most abundant
metal, have transitions that are detectable in X-rays.  However, even modern
X-ray instruments do not have the required resolution. \NeVI, \NeVIII\, \MgX\
and \OVI\ probe a similar, but typically cooler component of gas, but have much
longer transition wavelengths, e.g. $558 \units{\angstrom}$, $770
\units{\angstrom}$, $610 \units{\angstrom}$ and $1032 \units{\angstrom}$,
respectively. These are readily detectable with UV instruments and represent
good probes of hot, collisionally ionized gas, easier to detect than than their
X-ray counterparts, \OVII\ and \OVIII.

We now consider each of the ions considered in
Fig.~\ref{fig:column_density_radius}, one at a time. We begin with the column
density profile of \HI\, shown in the top row of
Fig.~\ref{fig:column_density_radius}. In general, the \HI\ column density
profile is similar at all redshifts in both simulations. The profile is
centrally peaked and falls off rapidly, typically decreasing by about $4$ orders
of magnitude within the first $100 \units{kpc}$, where it flattens to a near
constant number density of $10^{15} \units{cm^{-2}}$.

Between $z=3$ and $z=1$, the mean and scatter of the \HI\ column density at
fixed impact parameter agree well in both \apostle\ and \auriga. However, at
$z=0$, there is a slight systematic offset in \apostle\ where the \HI\ column
densities are about a factor of five higher than in \auriga. The offset
diminishes in both the centre, $<30 \units{kpc}$, and beyond $\rvir$.
The similarity of the \HI\ distributions in both simulations is not surprising.
In particular, we see in Fig.~\ref{fig:density_temperature_phase} that the
distribution of gas at high densities and low temperatures, where \HI\ typically
occurs, is very similar in the two (see Fig.~\ref{fig:contour_density}).

The second, third and fourth rows from the top of
Fig.~\ref{fig:column_density_radius} show the column density profiles of \OVI,
\NeVI\ and \NeVIII, respectively. We discuss the distributions of these ions
collectively, as they have similar general trends and typically probe gas at the
same density and temperature, as demonstrated in Fig.~\ref{fig:contour_density}.
\OVI, \NeVI\ and \NeVIII\ probe progressively hotter populations of dense gas,
increasing from $10^{5.5} \units{K}$ up to $10^{6} \units{K}$. \NeVI\ typically
probes a broader range of temperatures, $\sim 0.4 \units{dex}$, compared to
$\sim 0.2 \units{dex}$ for \OVI\ and \NeVIII. 

At $z=0$, the column densities of \OVI, \NeVI\ and \NeVIII\ in \auriga\ are
higher than in \apostle\ at all radii, but the difference is maximal at the
centre of the halo. Outside the central region, beyond $\sim 30 \units{kpc}$,
the differences in column density are fairly small, typically a factor of two or
less. Although the mean column density differs at a given impact parameter, the
range of column densities for the whole sample typically overlaps. The
differences, however, can be substantial at higher redshift, $z \geq 1$. In
Fig.~\ref{fig:column_density_radius} we see a general behaviour in the \OVI,
\NeVI\ and \NeVIII\ profiles in both \apostle\ and \auriga. In \apostle, the
column density distributions are relatively flat, typically decreasing by only
an order of magnitude over a radial range $3 \rvir$. In contrast, in \auriga,
the central column densities are typically higher and decrease much faster,
dropping by over four orders of magnitude over the same radial range.

The differences in the column density profiles of \OVI, \NeVI\ and \NeVIII\ at
$z \geq 1$ in \apostle\ and \auriga\ are most notable in both the innermost and
outermost regions. In \auriga, the column densities of these ions are higher in
the centre ($r < 50 \units{kpc}$), where they can be up to a hundred times
higher than in \apostle. However, as the column densities in \auriga\ decrease
so rapidly with increasing radius, the column densities in \apostle\ end up
being much higher in the outermost regions, $\sim 150 \units{kpc}$. In
particular, the column densities of these ions in the outer regions of \apostle\
are up to three orders of magnitude higher than in \auriga. We also note that
the column density variance at fixed impact parameter is much larger in \auriga\
than in \apostle, particularly in the outer regions.

The ions \OVII, \OVIII\ and \MgX\ probe some of the hottest gas surrounding our
galaxies. \OVII\ and \OVIII\ probe a broad range of hot gas, $T \sim 10^{5.5}
- 10^{6.5} \units{K}$, and $T \sim 10^{6} - 10^{6.5} \units{K}$, respectively.
\MgX\ probes gas in a narrower temperature range, $T \sim 10^{6} - 10^{6.25}
\units{K}$. 
These three ions, \OVII, \OVIII\ and \MgX, are shown in the fifth, sixth and
seventh rows from the top of Fig.~\ref{fig:column_density_radius}.  As with the
other ions, we find fairly good agreement between \apostle and \auriga at $z=0$,
with a considerable systematic offset for \MgX. Remarkably, the column densities
of \OVII\ and \OVIII\ agree to within $10\%$ beyond the inner $10 \units{kpc}$.
Both \apostle\ and \auriga\ predict relatively flat column densities as
a function of impact parameter for both \OVII\ and \OVIII.
The predictions for \MgX\ differ slightly. The column density of \MgX\ drops
rapidly beyond radius $\sim 200 \units{kpc}$ and flattens to a near-constant
value of $\sim 10^{12} \units{cm^{-2}}$ in both simulations. While the general
shape of the profiles are similar in both simulations, the column density in
\apostle\ is typically a factor of $2-5$ higher than in \auriga.

At $z\geq1$, \OVII, \OVIII\ and \MgX\ follow a similar trend as \OVI, \NeVI\ and
\NeVIII. In the central regions, the column densities in \auriga\ are either
higher (\OVII) or approximately equal to those in \apostle.  Further out, the
column densities are much higher in \apostle. This difference is due to the
steeply declining column densities in \auriga\ and the much flatter profiles in
\apostle.  The differences in \OVII, \OVIII\ and \MgX\ between \apostle\ and
\auriga\ are most prominent at $z=3$. \MgX\ is the most extreme case. In
\auriga\ it is only present within $40 \units{kpc}$ of the halo centre, whereas
in \apostle\ there are still very high \MgX column densities out to $150
\units{kpc}$ and even beyond.

At the present day, the column densities of all the ions considered in
Fig.~\ref{fig:column_density_radius} are broadly consistent in the two
simulations within the halo-to-halo scatter.
The main exception occurs in the central regions ($\leq 30 \units{kpc}$), where
the \auriga\ haloes typically have a peak in column density that can be up to
a factor of ten higher than in \apostle. These larger column densities at the
centres of \auriga\ haloes could be a signature of galactic fountains, which is
where enriched, hot outflows recycle within a small central region.
The other notable exception is the larger \MgX\ column density in \apostle, at
all radii, but particularly in the outermost regions. This enhancement likely
reflects the presence of more hot, $T \geq 10^6 \units{K}$, enriched gas at
large radii in \apostle\ arising from the larger spatial extent of the hot
outflowing material in this case.

We find a consistent trend among all the ions considered in
Fig.~\ref{fig:column_density_radius} for $z \geq 1$, with the exception of \HI.
This trend consists of higher column densities in the innermost regions of the
\auriga\ simulations, which then decline with impact parameter more rapidly than
in \apostle.  The important offshoot is that the \apostle\ haloes have
significantly higher column densities at large radii, $\geq 100 \units{kpc}$,
with the differences increasing with redshift.

The large column densities of \NeVI, \NeVIII, \OVII, \OVIII\ and \MgX\ in the
outer regions of the \apostle\ simulations at $z \sim 1 - 3$ are a strong
signature of hot, accretion-impeding outflows. A visual inspection of the
evolving temperature projections in Fig.~\ref{fig:temp_tracers} demonstrates
that hot gas, $T \sim 10^6 \units{K}$, in \apostle\ extends to radii of order
$\geq 4 \rvir$ by $z=3$. By contrast, the \auriga\ galaxies develop a much
cooler, $T \leq 10^{5.5}$, halo of gas which does not extend beyond $2 \rvir$.
This hot gas distribution in \auriga\ produces column densities that drop
rapidly at $\rvir \sim 50 \units{kpc}$ at $z=3$ and then drop even further
beyond $3 \rvir \approx 150 \units{kpc}$.

The peak in the column densities of \NeVI, \NeVIII, \OVI, \OVII\ and \OVIII\ at
the centre of the \auriga\ haloes can be readily understood by reference to the
density-temperature histograms in Fig.~\ref{fig:density_temperature_phase}.
Panels~(a) and~(b) show that there is a population of very dense gas, $n_{\rm H}
\geq 10^{-2} \units{cm^{-3}}$, with temperatures ranging between $T \approx
10^{4} - 10^{7} \units{K}$. As described in Section~\ref{sec:subgrid_diff}, this
gas component appears to be a product of a galactic fountain. The high density
of the gas leads to very short cooling times, $\sim 200 \units{Myr}$, so for
such a massive gas component to be present it must be continuously replenished
by the heating of dense gas. This gas is heated to $T \geq 10^6 \units{K}$ where
it cools at almost constant density before rejoining the ISM of the central
galaxy.  Therefore, the centrally-concentrated peak in ion column densities in
\auriga\ appears to be a strong signature of galactic fountains.

In summary, we find that the \apostle\ simulations produce almost flat column
density profiles for ions which probe hot gas, out to radii of $\sim 3 \rvir$
between $z=0-3$. These flat density profiles are produced by hot, outflowing gas
driven by SNe within the central galaxy. In contrast, the \auriga\ simulations
predict rapidly declining column densities with radius as the SNe driven
outflows are unable to eject large amounts of hot gas to such large radii.
Instead, the \auriga\ simulations generate galactic fountains where dense gas is
heated to high temperatures, $T \approx 10^7 \units{K}$, and then cools at
a high, almost constant, density.  This fountain produces an observable central
peak in the column densities of \NeVI, \NeVIII, \OVI, \OVII\ and \OVIII\ which
is not present in \apostle. 

\subsection{Dispersion measure}

\begin{figure*}
    \centering
    \includegraphics[width=1.00\textwidth]{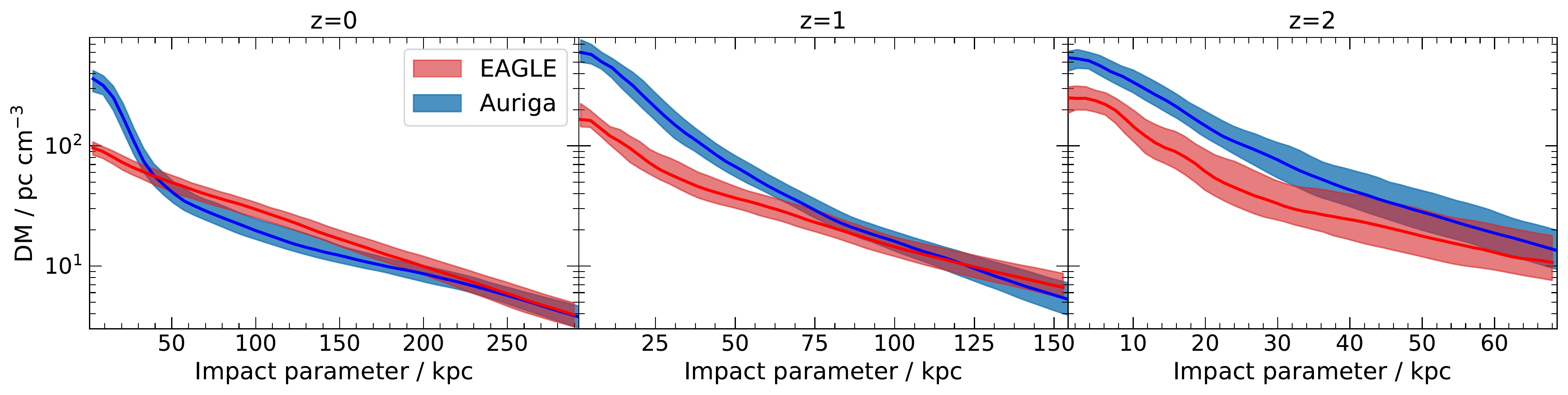}
    \caption{The dispersion measure of all four haloes in \apostle\ and
      \auriga\ at three redshifts, $z \sim 0,~1,~2$ from left to
      right, respectively.  The dispersion measure, at fixed radius,
      $r$, for each halo is calculated by taking the median dispersion
      measure of many sightlines through a small annulus of depth
      $\rvir$. We also compute the lower and upper quartiles of the
      dispersion measure in each annulus. The solid line shows the
      mean of the median dispersion measure, at each radius, for all
      four haloes. The shaded region illustrates the range between the
      mean of the lower and upper quartiles for all the haloes.  The
      dispersion measure at higher redshift is calculated in the frame
      of the halo at that redshift, not from an observer at the
      present day.  }
    \label{fig:dm_comparison}
\end{figure*}

The dispersion measure is a measure of the free electron column density along
a sightline and, potentially, one of the most useful metrics of the baryon
content of the Local Group. As discussed in Section~\ref{sec:calc_dm}, CHIME is
predicted to detect between $2$ and $40$ FRBs day$^{-1}$ over the whole sky
\citep{connor:2016}. The hot halo of M31 makes a significant contribution to the
dispersion measure of FRBs when their emission passes through the halo with an
impact parameter of $\leq 150 \units{kpc}$; this corresponds to an angle of
approximately $11 \degree$ on the sky, assuming that M31 is $\sim 770
\units{kpc}$ away from the MW. Therefore, M31 covers approximately $\sim 400
\units{deg^2}$, or $\sim 1\%$ of the sky.  This implies that the CHIME survey
should expect to detect roughly between $10$ and $150$ FRB's per year behind the
M31 halo. These can be compared with sightlines adjacent to M31. If the
foreground contribution from the MW is uniform, or at least smooth over a narrow
range of viewing angles, and the FRB population has the same redshift
distribution over this range, the differences in these dispersion measures will
be a direct reflection of the properties of the plasma in M31. The contribution
of M31 to the dispersion measure can then be used to infer the amount of hot gas
present in M31 and, thus, to constrain its baryon fraction.

In this section we compute the dispersion measure from several thousand random
sightlines through the four primary MW-like haloes at three redshifts,
$z=0,~1,~2$. The electron column density is calculated for parallel sightlines
as described in Section~\ref{sec:ions_obs}. These are projected directly through
$2\times \rvir$ of the primary halo at varying impact parameters. The dispersion
measure at a given radius of each halo is calculated by taking the median of
many sightlines in a small annulus. We project through the $x$, $y$ and $z$ axes
and combine the results. Additionally we compute the lower and upper quartiles
in each annulus.

The dispersion measure profiles in Fig.~\ref{fig:dm_comparison} represent
idealized observations of M31 from Earth, with contributions from the MW and
material beyond M31 removed (i.e. the IGM and other distant haloes). In
practice, we expect the contribution from halo gas in M31 to be large, and thus
to be readily detectable when compared with sightlines that do not pass through
M31. Although they are not realistic mocks of observations from Earth, the
results in Fig.~\ref{fig:dm_comparison} provide some insight into how the
dispersion measure of a halo varies with impact parameter and redshift, and thus
may help interpret observational data.  Later in this section, we discuss how
future work could improve the realism of the simulated profiles and how they
could be used to exploit the constraining power of future observations.

With current data it is only possible to make direct measurements of today's
dispersion measure around M31 and other Local Group. We do, however, include
results from higher redshifts in Fig.~\ref{fig:dm_comparison} to understand how
the dispersion measure evolves and to give insight on the possible background
contributions to observations. (The dispersion measure at higher redshift is
calculated in the frame of the halo at that redshift, not from an observer at
the present day.)

At the present day, both simulations predict a similar trend, with the
dispersion measure being highest in the central regions and declining with
increasing impact parameter. At $z=0$, the dispersion measure typically drops
from a peak value of $\geq 10^{2} \units{pc~cm^{-3}}$ in the centre of the halo
to $\sim 1 \units{pc~cm^{-3}}$ at $\rvir$, and continues to fall beyond this
radius (not shown in the figure).  Beyond the inner $50 \units{kpc}$, \apostle\
and \auriga\ predict similar profiles. 
The main difference occurs in the centre of the halo. In \apostle\ the
dispersion measure decreases with impact parameter at an almost constant rate.
In \auriga\ there is a peak at the very centre which drops rapidly out to $50
\units{kpc}$. Beyond that, both the slope and amplitude of the profiles in the
two simulations are approximately equal.

The peak in the dispersion measure in the inner regions of \auriga is also
present at higher redshifts. At $z \geq 1$ a similar peak, but of lower
amplitude, also appears in the \apostle simulations.
The peak at the centre of the \auriga\ haloes at all redshifts coincides with the
peak in the column densities of the ions that trace the warm-hot gas within the
CGM (see Fig.~\ref{fig:column_density_radius}) Thus, the origin of this large
dispersion measure is plausibly the same as that of the ions. Feedback produces
hot, metal-enriched, centrally-concentrated gas which is dense. This gas then
cools, at an almost constant density, before rejoining the ISM. The electron
mass will trace the gas mass in Fig.~\ref{fig:density_temperature_phase} (a). It
is the hot gas of atomic density $\geq 0.001 \units{cm^{-3}}$ in \auriga\ that
produces the centrally concentrated dispersion measure peak. This gas is not
present in \apostle\ at $z=0$ and, as a result, the profile is much flatter near
the centre.

At higher redshift, \auriga\ predicts a higher dispersion measure throughout the
halo. This behaviour is similar to that of the column densities of \NeVI,
\NeVIII, OVI and \OVII in the central regions seen in
Fig.~\ref{fig:column_density_radius}, which typically probe gas at temperature
$\sim 10^{5.5} \units{K}$. The difference in baryon mass in the haloes of
\apostle and \auriga drives the difference in the amplitude of the dispersion
measure profiles. 

At lower redshift, the baryon fraction of the \auriga\ haloes is still a factor
of two higher than the \apostle counterparts. However, the dispersion measure in
two simulations tends to agree reasonably well outside of the central region,
$\geq 50 \units{kpc}$. Inside the central region, the \auriga\ haloes boast
significantly higher dispersion measure, thus indicating that the extra baryonic
mass in \auriga, at present-day, is centrally concentrated. Efficient galactic
fountains in \auriga\ can continuously produce centrally concentrated hot gas.

In summary, the dispersion measure is a measure of the amount of ionised gas
along the line-of-sight and is strongly sensitive to the distribution of hot gas
around MW-mass haloes, which is mostly ionised. The similar dispersion measure
profiles in the outer regions of \apostle and \auriga at $z=0$ imply that the
haloes in the two simulations have similar amounts of hot gas in this region,
despite having large differences in baryon fraction. This is possible as the
extra baryons present in \auriga are centrally concentrated due to the galactic
fountains, which leads to a large central peak in the dispersion measure profile
in \auriga.

We predict that future surveys of dispersion measure inferred from FRBs should
be able to identify or exclude the existence of a galactic fountain in either
the MW or M31, through analysis of dispersion measure variation with impact
parameter within the central regions. We also expect that the background, e.g.
the contributions from the IGM and other intervening haloes at higher redshift,
should be larger if there are hot, spatially extended outflows at high redshift,
such as those found in \apostle. It may also be possible to identify the
presence of a hot galactic fountain by direct observation of X-ray emission
\citep{oppenheimer:2020b}.

In this analysis, we did not include material which is part of the ISM. When
calculating the free electron density we discarded gaseous material with an
atomic number density $\geq 0.1 \units{cm^{-1}}$ or a temperature $\leq 10^4
\units{K}$. Gas in this regime is not modelled explicitly in the simulations,
however the distribution and morphology of the cold gas is in reasonable
agreement with observations \citep{marinacci:2017}. The dispersion measure
profiles in the innermost regions may well be higher than predicted in this work
due to dispersion by ISM gas. However, predictions for the ISM suggest that it
contributes only $\leq 50 \units{pc~cm^{-3}}$ \citep{lorimer:2007}; thus the
central regions should be dominated by contributions from halo gas. 


Finally, we stress that realistic mock catalogues will be needed to interpret
future data. Constructing these will require combining high-resolution Local
Group simulations such as those presented here with large-volume cosmological
simulations to determine the expected background.

\subsection{Comparing to current observations}

In this section we compare our preliminary mock observations of the column
densities of \NeVIII for our sample of four MW-mass haloes with current
observations. The COS-Halos survey \citep{tumlinson:2013} and CASBaH survey
\citep{burchett:2019} are absorption line studies of galaxies in the UV. They
typically cover the redshift range $0.05 < z < 1.5$ and provide information on
column densities and covering fractions of \HI, \NeVIII\ and \OVI.
\cite{burchett:2019} collated a statistical sample of \NeVIII\ CGM absorbers.
This sample includes 29 CGM systems in the redshift range $z=0.5-1.5$, with
a median redshift, $z=0.68$, stellar masses in the range $10^{9.5}-10^{11.5}
\units{\msun}$ and impact parameters within $450 \units{kpc}$ of the central
galaxy.

\begin{figure}
    \centering
    \includegraphics[width=1.00\linewidth]{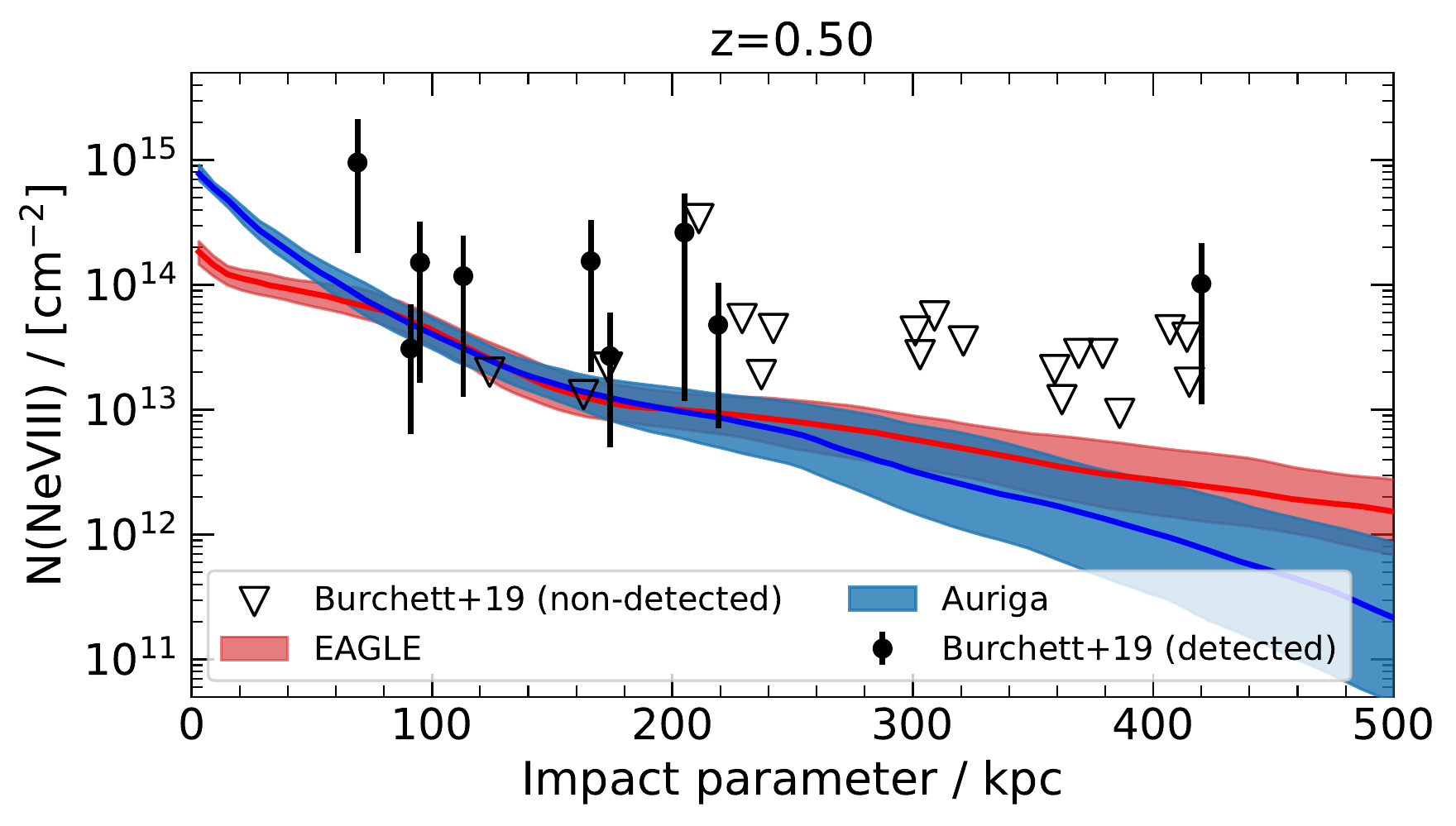}
    \caption{The atomic column number density of \NeVIII\ as a
      function of impact parameter from the centre of the primary halo
      in both the \apostle\ (red) and \auriga\ (blue) simulations out
      to $500 \units{kpc}$ at $z=0.5$.  The column density at a fixed
      radius, $r$, for a single halo is calculated by taking the median
      column density of many sightlines through a small annulus. We
      also compute the lower and upper quartiles of the column density
      in each annulus. The solid line shows the mean of the median
      column density, at each radius, for all four haloes. The shaded
      region illustrates the range between the mean of the lower, and
      upper quartiles, for all of the haloes sampled.  We also include
      the observational detections (solid black circles) and
      non-detections (empty triangles) of
      \protect\cite{burchett:2019}.}
    \label{fig:burchett_col_density}
\end{figure}

In Fig.~\ref{fig:burchett_col_density} we compare the column density of \NeVIII,
as in Fig.~\ref{fig:column_density_radius}, at $z=0.5$ for both \apostle\ and
\auriga, with observational data, including both the detections (solid black
circles) and non-detections (empty triangles) of \cite{burchett:2019}. The
highest inferred column density of \NeVIII\ is $14.98\pm0.09 \units{cm^{-2}}$ at
an impact parameter of $69 \units{kpc}$ and redshift, $z=0.93$. The central
galaxy of this system has an estimated stellar mass of $10^{11.2}
\units{\msun}$, slightly larger than our simulated galaxies. This high observed
column density is larger than found in any of the predictions of \apostle\ and
\auriga, as seen in the figure. However, column densities this high are not
uncommon at lower impact parameters in \auriga.

The column densities at slightly larger radii, $100-200 \units{kpc}$, are
consistent with the predictions of both simulations, with almost all of the
observational detections in this range overlapping the results from our
simulations within the uncertainties. In the outer regions, the observations are
dominated by upper limits which are higher than, and thus consistent with the
inferred column densities in the simulations.

The observations follow the general trend that the inner regions are dominated
by detections of $\approx 10^{14} \units{cm^{-2}}$, whereas the outer regions
are mostly upper limits between $\approx 10^{13.5} - 10^{14} \units{cm^{-2}}$.
This is suggestive of a \NeVIII\ column density profile which typically declines
by $\geq 0.5 \units{dex}$ between an impact parameter of $150 \units{kpc}$ and
$300 \units{kpc}$. This is also seen in both \apostle and \auriga. \apostle\
better recovers the (approximately) flat distribution of \NeVIII\ detections;
however, \auriga\ agrees better with the higher central column densities. In any
case, the model preferences are driven by two data points, the ones with the
lowest and highest impact parameters. Therefore the model choice is subjective,
and there is no clear preference towards either \apostle\ or \auriga.

\section{Discussion and Conclusions}\label{sec:conclusions}

The two simulations that we have analysed in this work have been shown to
reproduce many galaxy observables even though they involve different galaxy
formation models and hydrodynamical schemes. In particular the large-volume
\apostle\ simulations reproduce the galaxy stellar mass function
\citep{schaye:2015}, the evolution of galaxy masses \citep{furlong:2015}, sizes
\citep{furlong:2017} and colours \citep{trayford:2015}. Similarly, large volume
simulations with a similar model to \auriga\ have successfully reproduced the
scaling relations and evolution of galaxy sizes \citep{genel:2018}, the
formation of realistic disc galaxies \citep{pillepich:2019}, the gas-phase
mass-metallicity relation \citep{torrey:2019} and the diversity of kinematic
properties observed in the MW-type galaxies \citep{lovell:2018}.

In this work, we have analysed the emergent baryon cycle around two Local
Group-like volumes centred around a pair of haloes similar to those of the MW and
M31. We investigated how the baryon cycle differed when using the different
subgrid models of the \apostle and \auriga\ simulations. While these models are
similar, they have significantly different implementations of SNe feedback.
\apostle\ injects all the energy from SNe in the form of a thermal energy
`dump', whereas \auriga\ uses hydrodynamically decoupled `wind' particles that
carry mass, energy, momentum and metals away from the ISM to lower density
regions of the galactic halo.

In Section~\ref{sec:baryon_evolution}, we explored the effects of the different
feedback implementations on baryonic evolution, particularly the baryon
fraction, in and around the two primary haloes as a function of time. We found
the minimum baryon fraction within a sphere around a primary \apostle\ galaxy to
be $\approx 40\%$ of the cosmic baryon budget, which is approximately half the
value found in \auriga. Furthermore, the \apostle simulations exhibit a baryon
deficiency of $\geq 10\%$ within a radius  $\geq 1 \units{Mpc}$ (comoving) of
the halo, extending to $\approx 2 \units{Mpc}$ at the present day. Thus, in
\apostle, the Local Group is a baryon deficient environment. Conversely, in the
\auriga\ simulations the baryon fraction is within $5\%$ of unity at all radii
$\geq 0.5 \units{Mpc}$ (comoving), and at all redshifts. This difference in the
baryon evolution is remarkable given that both simulations use the exact same
initial conditions and produce central galaxies with relatively similar stellar
properties. This is consistent with the findings of \cite{mitchell:2021} which
show the gas mass, and thus density, of the CGM are more sensitive to the baryon
cycle than is the case for the properties of the central galaxy.

In Section~\ref{sec:missing_halo_baryons} we conducted a census of all the
baryons expected to lie within $\rvir$ at the present day due to gravitational
forces alone (which we called `predestined'). In \apostle\ we found that $\sim
35\%$ of the baryonic counterparts of the $z=0$ dark matter halo particles
inhabit the primary halo, whereas in \auriga\ approximately $70\%$ do.
Furthermore, in \apostle\ we found that almost half of the baryon counterparts
of dark matter particles that are missing \textit{never entered the halo}: they
are `impeded'. By contrast, in \auriga\ almost $ 90\%$ of the absent baryons
entered the halo before being ejected.

We also found that the physical extent of ejected and impeded baryons, in both
\apostle and \auriga\ is such that there is baryonic mixing between the two
primary haloes.  This baryonic mass transfer, shown in
Fig.~\ref{fig:missing_halo_baryons_projected}, indicates that the presence of
M31 may influence the evolution of the MW and viceversa
\citep[see][]{borrow:2020}.

The large `impeded' gas component in \apostle is produced by halo-scale fast
outflows with high covering fractions. The \auriga simulations do not produce
sufficiently large outflows to impede accretion significantly.
However, SNe feedback in both, \apostle and \auriga\ inject a similar amount of
energy per unit mass into the surrounding gas. Thus, the different fates of the
energetic gas must result from the method of injecting the energy, or the
subsequent evolution of the gas (followed with two different hydrodynamical
scheme.). We strongly suspect that it is the former that is responsible for the
different outcomes.

This differences in the SNe subgrid prescriptions in \apostle and \auriga\
result in the post-SNe temperature of the gas in \apostle being much higher that
in \auriga\ and, crucially, greater than $10^{7} \units{K}$. Thus, in \apostle
radiative cooling is inefficient and the gas expands adiabatically. This
expansion produces hot, low-density parcels of gas which are buoyant and
accelerate outwards through the halo \citep{bower:2018}. Conversely, in \auriga\
the lower post-SNe temperature makes radiatively cooling more efficient and the
gas cools rapidly, at almost constant density, and recycles on a short
timescale.

To summarise, we have found that the processes that regulate the rate of star
formation in a MW-mass galaxy can be classified into two broad categories:
ejective and preventative.
The \auriga\ simulations are dominated by ejective feedback with relatively
short recycling times, regulating star formation without significantly reducing
the halo baryon fraction.
However, in \apostle feedback at high redshift ejects a large mass of gas beyond
$\rvir$. This gas has such a large covering fraction and outward pressure that
it can suppress the cosmological accretion of gas. These findings are consistent
with the results from the large volume \apostle\ simulations
\citep{mitchell:2019, mitchell:2020b} and the NIHAO simulations
\citep{Tollet:2019} which all identify reduced cosmic gas accretion rates due to
feedback processes.
The FIRE simulations \citep{muratov:2015} present some evidence that bursts of
star formation at high redshift can suppress gas accretion into the
\textit{inner} halo; however, it is not clear if this suppression extends to
$\rvir$ or low redshift. The FIRE simulations also produce (almost) baryonically
closed MW-mass haloes \citep{hafen:2019}, thus more closely resembling those in
\auriga, than \apostle.
These results highlight a fundamental difference in the outcome of the feedback
processes in various subgrid implementations.

A caveat is that the stellar mass of the central galaxies in \apostle\ is
typically a factor of two lower than in \auriga. In principle, we could `tune'
the feedback parameters in \apostle to produce more massive galaxies in $10^{12}
\units{\msun}$ haloes and still end up with baryon-deficient haloes due to
preventative feedback. \cite{schaller:2015b} demonstrates that the `weak SNe'
\apostle\ variant can produce MW-mass haloes with stellar masses consistent with
the \auriga galaxies, albeit with weaker feedback than used in \auriga.
Furthermore, \cite{wright:2020} show that the rate of gas accretion at $\rvir$
for the same `weak SNe' simulation differs from the \apostle\ reference model
used in this work by less than $20\%$. Thus, the \apostle\ `weak SNe' simulation
produces \auriga-mass galaxies, while still suppressing cosmological gas
accretion as shown in this work.

While we identify SNe feedback as the leading cause of differences in the baryon
cycle around our simulated galaxies, there are other differences in the two
simulations that could contribute.
Namely, both simulations use different hydrodynamical solvers; however
\cite{schaller:2015b} and \cite{hopkins:2018} have demonstrated that the
treatment of feedback is more significant in determining the outcome of
simulations of this kind than the details of the hydrodynamics solver.
\apostle\ and \auriga\ also use different implementations of AGN feedback.
However, there is evidence to suggest that the effects of these AGN models on
the gaseous baryon cycle are subdominant for haloes of mass $\sim 10^{12}
\units{\msun}$.
\cite{voort:2021} shows the total gas mass within the virial radius of a MW-mass
halo simulated with the \auriga\ model both with and without AGN feedback differ
by only a few percent. However, the inclusion of AGN feedback decreases the $z=0$
stellar mass by $\sim 30\%$.
\cite{davies:2020} show that AGN feedback in the \apostle\ reference model injects four times less energy than SNe in $\sim 10^{12} \units{\msun}$ haloes. Furthermore, Figure~11 of \cite{crain:2015} shows the growth of black holes in the `ViscLo` AGN model in \apostle, used in this work, is significantly reduced relative to the reference model. As a result of this, the present-day black hole masses in the \apostle\ haloes presented in this work are all $\leq 10^{6.8} \units{\msun}$ and therefore appear to have no significant effect on the baryon fraction of the haloes \citep{bower:2017}. However, \cite{davies:2020} and \cite{oppenheimer:2020a} demonstrate that AGN feedback can significantly reduce the baryon fraction of MW-mass haloes which host black holes of mass $\geq 10^{7.0-7.3} \units{\msun}$.

In Section~\ref{sec:observables} we investigated the observable signatures of
strong, accretion-impeding outflows in \apostle\ and efficient galactic
fountains in \auriga. We concluded that the \apostle\ simulations produce almost
flat column density profiles of ions which probe hot gas --\NeVI, \NeVIII, \OVI,
\OVII, \OVIII and \MgX-- out to radii $\sim 3 \rvir$ between $z=0-3$. These flat
profiles are a signature of hot, outflowing gas driven by SNe within the central
galaxy. In contrast, the \auriga\ simulations predict rapidly declining column
densities with radius, as the SNe driven outflows are unable to eject large
amounts of hot gas to such large radii. We attempted to constrain the two
subgrid models by comparing to the data on \NeVIII presented by
\cite{burchett:2019}. Unfortunately, these data are not constraining due to the
small number of detections at very small and very large impact parameters.

We also investigated the dispersion measure, which probes the integrated free
electron density along the line-of-sight. We found that the main difference
between the dispersion measure profiles in \auriga\ and \apostle\ is also at the
very centre. In \auriga, there is a peak at impact parameter $\leq 50
\units{kpc}$, not found in \apostle. The dispersion measure is
a promising observational diagnostic of the evolution of baryons around
galaxies. A combination of high-resolution Local Group analogues, like those
presented in this work, and large-volume cosmological simulations would
facilitate the production of more realistic mock catalogues of FRBs. Large
cosmological volumes allow modelling random sightlines out to the typical FRB
redshifts, $z\sim 2$. These background contributions can be added to predictions
from Local Group analogues to make realistic mocks of what would be seen from
the Earth when looking, for example, in the direction of M31. These can be
compared with real sightlines in the direction of M31 and just adjacent to it.
As future surveys should detect hundreds of FRBs per year behind M31, these
observations should be able to constrain models and shed light on the dominant
processes involved in the galactic baryon cycle.

To conclude, we find that ejective and preventative feedback work in tandem to
reduce the amount of gas within haloes in the \apostle galaxy formation model. In
\auriga, MW-mass galactic haloes are almost `baryonically closed', as ejective
feedback beyond $\rvir$ typically re-accretes and does not significantly impede
further cosmic gas accretion. Future observations of FRBs and CGM ion absorption
should provide valuable data to compare and constrain different galaxy formation
models.
Dwarf galaxies may also provide a suitable laboratory for studying the baryon
cycle. In particular, Fig.~\ref{fig:stellar_halo} shows there are significant
differences in the stellar-mass/halo-mass relation of lower mass galaxies in the
two models; these differences carry over into the baryon fraction of these
objects. It is also likely that the size of the heated gas coronae around haloes
within the Local Group will affect the number of star-free dark matter haloes,
which are even less massive than dwarf galaxies, in the local environment
\citep{Benitez-LLambay:2017,sykes:2019}.

\section*{Acknowledgements}

We thank Joop Schaye and Josh Borrow for discussions and comments, which significantly improved the manuscript.
This work was supported by the Science and Technology Facilities Council (STFC)
consolidated grant ST/P000541/1. AJK acknowledges an STFC studentship grant
ST/S505365/1. CSF acknowledges support by the European Research Council (ERC)
through Advanced Investigator grant DMIDAS (GA 786910). AF acknowledges support by a UKRI Future Leaders Fellowship (grant no MR/T042362/1) and Leverhulme Trust.
This work used the
DiRAC@Durham facility managed by the Institute for Computational Cosmology on
behalf of the STFC DiRAC HPC Facility (www.dirac.ac.uk). The equipment was
funded by BEIS capital funding via STFC capital grants ST/K00042X/1,
ST/P002293/1, ST/R002371/1 and ST/S002502/1, Durham University and STFC
operations grant ST/R000832/1. DiRAC is part of the National e-Infrastructure. 

\section*{Software Citations}

This paper made use of the following software packages:
\begin{itemize}
    \item \textsc{gadget} \citep{springel:2005}
    \item Arepo \citep{springel:2010}
    \item {\tt python} \citep{python}, with the libraries:
    \begin{itemize}
        \item {\tt numpy} \citep{numpy}
        \item {\tt scipy} \citep{scipy}
        \item {\tt h5py} \citep{hdf5}
        \item {\tt matplotlib} \citep{mpl}
        \item {\tt numba} \citep{numba}
        \item {\tt unyt} \citep{unyt}
    \end{itemize}
\end{itemize}

\bibliographystyle{mnras}
\bibliography{ref}

\appendix

\section{Stellar disc properties}\label{app:stellar_properties}

\begin{figure*}
    \centering
    \includegraphics[width=\linewidth]{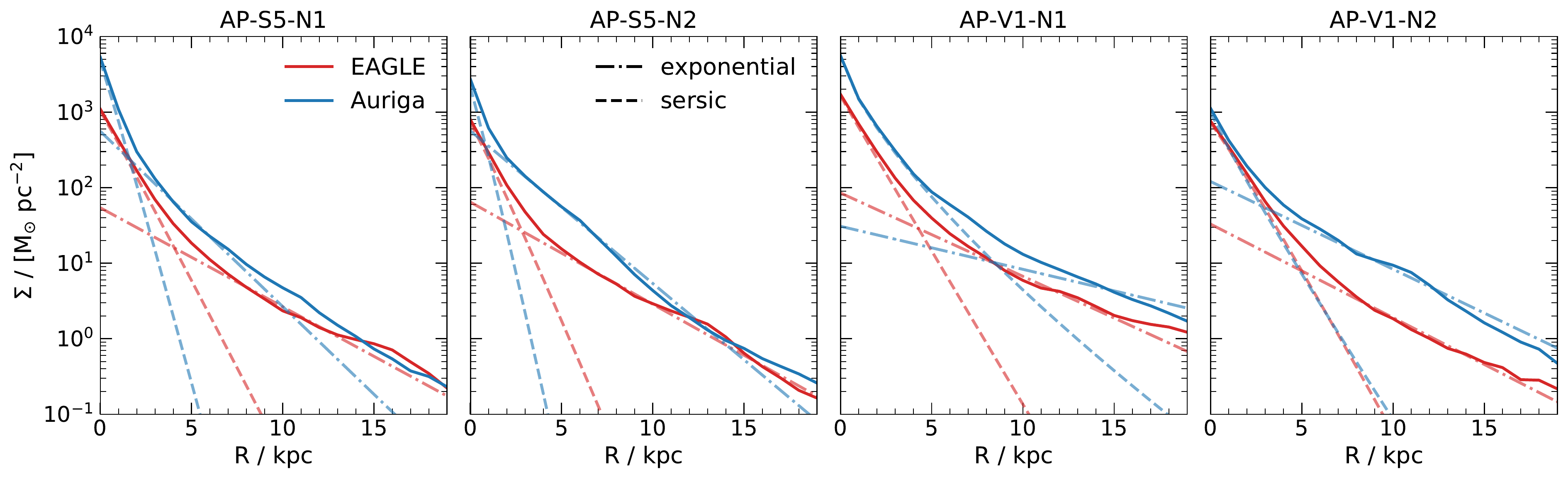}
    \caption{The radial stellar surface density profiles of the four
      primary haloes. The surface densities are calculated using all
      the stellar particles within a physical height of
      $\pm 5 \units{kpc}$ from the galactic plane; these are shown as 
      solid lines.  The radial profiles are simultaneously fit with a
      Sersic (dashed) and exponential (dashed-dotted) profile using a
      non-linear least-squares method. Results for the \apostle\ and
      \auriga\ simulations are shown in red and blue 
      respectively. The best-fit parameters may be found in
      Table~\protect\ref{tab:sim_properties}.  }
    \label{fig:surface_den}
\end{figure*}

Fig. \ref{fig:surface_den} shows stellar surface density profiles, for stellar
mass within $\pm 5 \units{kpc}$ of the mid plane in the vertical direction, for
all simulations at $z=0$. The profiles are simultaneously fit with a linear sum
of an exponential profile of scale radius, $R_{\rm{D}}$, and a Sersic profile of
the form $\exp (R/R_{\rm{eff}})^n$ \citep{sersic:1963}. The best fit values are
calculated using the least squares method to logarithm density profile.

\bsp
\label{lastpage}
\end{document}